\documentclass[traditabstract]{aa}
\usepackage[varg]{txfonts}
\usepackage{graphicx}
\usepackage{txfonts}
\usepackage{multirow}
\usepackage{graphicx}
\usepackage{txfonts}
\usepackage{textcomp}
\usepackage{natbib}
\bibpunct{(}{)}{;}{a}{}{,} 
\usepackage{txfonts}
\usepackage{rotating}
\usepackage{amssymb}
\begin{document}

   \title{The Gaia-ESO Survey: Kinematics of seven Galactic globular clusters\thanks{
 Based on data products from observations made with ESO telescopes at the La Silla Paranal Observatory under programme 188.B-3002 (the public Gaia-ESO spectroscopic survey, 
 PIs Gilmore \& Randich) and on the archive data of the programmes 62.N-0236,  
63.L-0439,  
65.L-0561,  
68.D-0212,  
68.D-0265,  
69.D-0582,  
064.L-0255,  
065.L-0463,  
071.D-0205,
073.D-0211,
073.D-0695,
075.D-0492,
077.D-0246,
077.D-0652,
079.D-0645,
080.B-0489, 
080.D-0106,  
081.D-0253,  
082.B-0386,  
083.B-0083,  
083.D-0208,  
083.D-0798,  
085.D-0205, 
086.D-0141,  
088.A-9012,  
088.B-0403,  
088.B-0492,  
088.D-0026,  
088.D-0519,  
089.D-0038,  
164.O-0561,  
386.D-0086. 
 }}

\author{C. Lardo\inst{1,2},
   E. Pancino\inst{2,3} , M. Bellazzini\inst{2} , A. Bragaglia\inst{2} , P. Donati\inst{2},  
   G. Gilmore\inst{4}, S. Randich\inst{5}, S. Feltzing\inst{6}, R.~D. Jeffries\inst{7} ,         
   A. Vallenari\inst{8}, E.~J. Alfaro\inst{9} , C. Allende Prieto\inst{10,11}, E. Flaccomio\inst{12},          
  S.~E. Koposov\inst{4,13}, A. Recio-Blanco\inst{14}, M. Bergemann\inst{4}, G. Carraro\inst{15},               
  M.~T. Costado\inst{9}, F. Damiani\inst{12},  A. Hourihane\inst{4}, P. Jofr\'e\inst{4}, P. de Laverny\inst{14} ,            
G. Marconi\inst{15}, T. Masseron\inst{4}, L. Morbidelli\inst{5}, G.~G. Sacco\inst{5} ,C.~C. Worley\inst{4} }

 \authorrunning{Lardo et al.}
  
  \institute{ Astrophysics Research Institute, Liverpool John Moores University, 146 Brownlow Hill, Liverpool L3 5RF, United Kingdom
   \email{C.Lardo@ljmu.ac.uk}
   \and 
   INAF - Osservatorio Astronomico di Bologna, via Ranzani 1, 40127, Bologna, Italy
   \and
   ASI Science Data Center, Via del Politecnico SNC, 00133 Roma, Italy
    \and 
   Institute of Astronomy, University of Cambridge, Madingley Road, Cambridge CB3 0HA, United Kingdom
   \and
   INAF - Osservatorio Astrofisico di Arcetri, Largo E. Fermi 5, 50125, Florence, Italy
   \and
   Lund Observatory, Department of Astronomy and Theoretical Physics, Box 43, SE-221 00 Lund, Sweden
   \and
    Astrophysics Group, Research Institute for the Environment, Physical Sciences and Applied Mathematics, Keele University, Keele, Staffordshire ST5 5BG, United Kingdom   
    \and
    INAF - Padova Observatory, Vicolo dell'Osservatorio 5, 35122 Padova, Italy
    \and
    Instituto de Astrof\'{i}sica de Andaluc\'{i}a-CSIC, Apdo. 3004, 18080, Granada, Spain
    \and
    Instituto de Astrof\'{\i}sica de Canarias, E-38205 La Laguna, Tenerife, Spain
    \and
    Universidad de La Laguna, Dept. Astrof\'{\i}sica, E-38206 La Laguna, Tenerife, Spain
    \and
    INAF - Osservatorio Astronomico di Palermo, Piazza del Parlamento 1, 90134, Palermo, Italy
    \and
    Moscow MV Lomonosov State University, Sternberg Astronomical Institute, Moscow 119992, Russia
    \and
    Laboratoire Lagrange (UMR7293), Universit\'e de Nice Sophia Antipolis, CNRS,Observatoire de la C\^ote d'Azur, CS 34229,F-06304 Nice cedex 4, France
    \and
    European Southern Observatory, Alonso de Cordova 3107 Vitacura, Santiago de Chile, Chile
}

   \date{Received XXX  XX, XXXX; accepted XXX XX, XXXX}

  \abstract{The Gaia-ESO survey is a large public spectroscopic survey aimed at investigating 
   the origin and formation history of our Galaxy by collecting 
   spectroscopy  of representative samples (about 10$^{5}$ Milky Way stars) of
   all Galactic stellar populations, in the field and in clusters.
   The survey uses globular clusters as intra- and inter-survey 
   calibrators, deriving stellar atmospheric parameters and abundances of a significant number 
   of stars in clusters, along with radial velocity determinations.
   We used precise radial velocities  of a large number of stars in seven globular clusters (NGC 1851, NGC 2808, NGC 4372, NGC 4833, NGC 5927, NGC 6752, and 
   NGC 7078) to validate pipeline results and to preliminarily investigate the cluster internal kinematics. 
   Radial velocity measurements were extracted from FLAMES/GIRAFFE spectra 
   processed by the survey pipeline as part of the second internal data release of data products to ESO. 
   We complemented our sample with ESO archival data obtained with different
   instrument configurations. Reliable radial velocity measurements 
   for 1513 bona fide cluster star members were obtained in total.
   We measured systemic rotation, estimated central velocity dispersions, and present velocity dispersion profiles
   of all the selected clusters, providing the first velocity dispersion curve and the first estimate of the central velocity dispersion
   for the cluster NGC~5927.
   Finally, we explore the possible link between cluster kinematics and other physical parameters. The analysis we present here demonstrates that Gaia-ESO survey data 
   are sufficiently accurate to be used in studies of kinematics of
   stellar systems and stellar populations in the Milky Way.}


   \maketitle
%

\section{Introduction}\label{INTRODUZIONE}

Globular clusters (GCs) have always been regarded as unique 
laboratories to explore many aspects of stellar dynamics \citep{meylan97}.
In a first approximation, they can  be considered spherically symmetric, 
non-rotating, and isotropic; but, as improved observations and new theoretical studies 
have become available, it became clear that they are complex 
(see \citealt{zocchi12}, \citealt{bianchini13}, and \citealt{kacharov14} for a discussion).
In particular, radial anisotropy \citep{ibata13}, deviations from sphericity \citep{white87,chen10},
mass segregation \citep{dacosta82}, signatures of core-collapse \citep{djorgovski84}, and velocity 
dispersion inflated by unresolved binary stars \citep{bradford11} have been observed 
and need to be explained in the framework of a dynamical scenario. 

Different physical mechanisms may determine these deviations from the perfect sphere:
velocity anisotropies, tidal stresses, and internal rotation \citep{goodwin97,gnedin99,vandenbergh08,bianchini13,kacharov14}.
The idea that internal rotation plays a fundamental part in determining the morphology of GCs was formulated some 50 years ago \citep{king61}.
Internal rotation has been detected in a growing number of 
GCs from line-of-sight velocity measurements (see, e.g., \citealp{bellazzini12}, hereafter B12)
and, in a few cases, from proper motion measurements 
(e.g., \citealp{vanleeuwen00}; \citealp{anderson03}).
The interest in the GC internal rotation is manifold.
Analytical \citep{longaretti97}, Fokker-Planck \citep{spurzem99}, and N-body models \citep{ernst07}
demonstrated that an overall (differential) rotation has a noticeable influence 
on stellar systems that evolve by two-body relaxation. 
In particular, it accelerates the core-collapse time
scales \citep{ernst07}\footnote{This effect seems to vanish 
for isolated two-mass N-body models \citep{ernst07}.}.
Internal rotation may also play an indirect role in the open question 
of whether there are intermediate-mass black holes (IMBH) in some GCs.
In fact, the detection of strong gradients in the velocity dispersion profile toward the cluster core 
is often interpreted as a hint of the presence of an IMBH \citep{baumgardt05}, but 
the evidence gathered so far in support of the existence of IMBHs is inconclusive and controversial, and
none of the published studies \citep{vandermarel10,lutzgendorf11,lanzoni13}
did consider differential rotation, which, together with anisotropy, can yield gradients in the velocity dispersion profiles \citep{varri12,bianchini13}.
Finally, recent investigations indicate that rotation could be a key ingredient 
in the formation of multiple generations of stars in GCs \citep{bekki10,mastrobuono13}.

In this science verification paper, we make use of the Gaia-ESO survey radial velocity 
determination to perform a kinematic analysis for seven Galactic GCs (NGC 1851, NGC 2808, NGC 4372, NGC 4833, NGC 5927, NGC 6752, and
M15),
following the same scheme as B12.
The samples we analyse were collected for a completely different scientific purpose, therefore  
they present intrinsic limitations for the characterisation of the cluster kinematics. 
The most recent dedicated studies used up to several hundred radial
velocity determinations (see e.g., \citealp{lane09,lane10a,lane10b}),
in some cases complemented with proper motions \citep{vandeven06,vandenbosch06,mclaughlin06,watkins13},
while we have V$_{\rm{r}}$ determinations for fewer than 100 stars 
for some clusters (i.e., NGC 2808, NGC 4833, NGC 5927).
Furthermore, the cluster members are unevenly distributed with radius 
within each cluster, with the large majority of the stars lying at distances greater than the half-light radius,
because it is difficult to allocate fibers in the very crowded central regions.

Still, our analysis {\em (a)} provides a validation of the Gaia-ESO survey radial velocities in a controlled sample, {\em (b)}
provides (and makes publicly available) additional observational material to study the kinematics of the considered 
clusters, and {\em (c)} at least in one case, NGC~5927, provides the first insight into the cluster kinematics.

This paper is structured as follows:
We begin by describing the data and the membership 
selection for each cluster in Sect.~\ref{DATASET}. 
We compute systemic velocities and velocity dispersions in Sect.~\ref{sigma}, 
as well as rotations (in Sect.~\ref{rotazioni}). In Sect.~\ref{parametri} we investigate 
the links between kinematics and cluster parameters.
Finally, our concluding remarks are presented in Sect.~\ref{Conclusioni}.

\begin{table*}
\footnotesize
 \caption{Archive spectra inventory. Note that the number of spectra quoted here is the total number of spectra obtained before Galactic contaminants were removed. }
 \renewcommand{\tabcolsep}{0.22cm}
 \label{ARCHIVE}

 \centering\begin{tabular}{l l  c c c c c c c c c c c c c c c }
            \hline\hline
          &          & HR4   &  HR9A  &   HR9B  &   HR10   &  HR11   &    HR13  &   HR14A  & HR14B  &  HR15N &  HR19A &   HR21   \\    
Target    & Type     & 427.2 &  525.8 &   525.8 &   548.8  &  572.8  &    627.3 &   651.5  & 651.5  &  665.0 &  805.3 &   875.7  \\ 
\hline
M 15      & Archive  &       &        &         &          &   83    &     155  &    81    &       &         &        &          \\  
          & Gaia-ESO &       &        &         &     79   &         &   &          &          &         &        &     80   \\ 
NGC 1851  & Archive  &       &  104   &   204   &          & 105     &    196        &          &          &         &   83   &         \\ 
          & Gaia-ESO &       &        &         &     94   &         &   &          &          &         &        &     92   \\ 
NGC 2808  & Archive  &       &        &         &          &         &      113 &     113  &           &         &    120 &          \\    
          & Gaia-ESO &       &        &         &     65   &         &   &          &          &         &        &     63   \\    
NGC 4372  & Archive  &       &        &         &          &         &   234     &    122   &          &         &        &          \\  
          & Gaia-ESO &       &        &         &     103  &         &   &          &          &         &        &     103  \\    
NGC 4833  & Archive  &       &        &         &          &   112   &  114  &          &          &         &        &          \\  
          & Gaia-ESO &       &        &         &    81    &         &   &          &          &         &        &      81  \\   
NGC 5927  & Gaia-ESO &       &        &         &    110   &         &   &          &          &         &        &     110  \\    
NGC 6752  & Archive  &   121 &        &     99  &     100  &    429  &      515 &      &     99   &    233  &        &     231  \\    
          & Gaia-ESO &       &        &         &     108  &         &   &          &          &         &        &     108  \\    
  \hline        
            
           \end{tabular}

\end{table*}

\section{Sample and radial velocity measurements}\label{DATASET}
\subsection{Data sets}

The Gaia-ESO Survey is a public spectroscopic survey that
uses the FLAMES multi-object spectrograph on the VLT 
UT-2 (Kueyen) telescope to obtain high-quality, uniformly calibrated 
spectroscopy of 100 000 stars in the Milky Way (\citealp{gilmoremsn}, \citealp{randichmsn}).
The survey targets stars in the halo, bulge, thick and thin discs, and in star-forming regions and open clusters to characterize
the chemistry and kinematics of these populations. 
When combined with precise astrometry from the
recently launched Gaia satellite \citep{perryman01}, the enormous dataset
will provide three-dimensional spatial distribution and kinematics, stellar parameters,
and chemical abundances for a significant number of stars 
in the Galaxy.

In addition to the main targets, the Gaia-ESO survey is observing GCs
as intra- and inter-survey astrophysical calibrators, 
deriving stellar atmospheric parameters, abundances, and radial velocities (V$_{\rm{r}}$) 
for typically a hundred red giant branch (RGB) stars in each cluster.
GCs  were selected among those used by other surveys as RAVE 
\citep{steinmetz06, zwitter08,siebert11,lane11}, GALAH \citep{zucker13}, and APOGEE \citep{frinchaboy12, frinchaboy13a,frinchaboy13b, meszaros13} where possible.
The photometric catalogues for the selected clusters are generally based on UBVI archival images, 
collected at the Wide-Field Imager (WFI) at the 2.2-m ESO-MPI telescope. The WFI covers a total field of view of 34\arcmin $\times$   33\arcmin, 
consisting of 8, 2048 $\times$ 4096 EEV-CCDs with a pixel size of 0.238\arcsec . These images were pre-reduced
using  the IRAF package MSCRED \citep{valdes98}, while the stellar photometry was derived by using the DAOPHOT II and ALLSTAR programs
 \citep{stetson87,stetson92}. Details on the preproduction, calibration, and full photometric catalogues will be published elsewhere.
We thus created the initial sample that includes as many clusters as possible from the other surveys, and filled in the gaps in [Fe/H] with clusters visible from the South that have public photometry 
data. To select the targets within each cluster, we generally observed RGB stars and performed a survey of FLAMES data in the ESO archive and in the literature (when available) to select probable members. To maximise our chances of obtaining reliable parameters for GC, we gave highest priority to GIRAFFE targets that already had archival observations in different setups and avoided repeating stars that already had UVES observations in the Gaia-ESO survey setups.
Additional details of the cluster selection criteria and observational strategy will be given in a 
forthcoming paper (Pancino et al., in preparation).

Our sample consists of seven Galactic GCs
observed by the Gaia-ESO survey.
The observations were performed between December 2011 and 
September 2013 and consist of one pointing for each GC, using 
the two FLAMES-GIRAFFE\footnote{We considered only stars observed with GIRAFFE to preserve homogeneity.}
setups that are used to observe the main field targets of the survey
\citep{gilmoremsn,randichmsn}: the high-resolution setups HR 10 (centred on 5488\AA, with a 
spectral resolution R=19800) and HR 21 (centred on 8757\AA, with a spectral resolution R=16200).

As the Gaia-ESO survey is a public ESO spectroscopic survey, raw spectra are available 
in the ESO archive\footnote{\url{http://archive.eso.org/eso/eso_archive_main.html}} as 
soon as targets are observed. 
Pipeline-reduced spectra for a fraction of the target stars observed in the first six months of 
observations are already available at the address \url{http://archive.eso.org/wdb/wdb/adp/phase3_main/form}.
Part of the data analysed in this paper are included in the second internal release and will become 
public within a few months.
In addition to the Gaia-ESO survey spectra, we complement our dataset with archive FLAMES data observed 
with different instrumental configurations\footnote{See \url{http://www.eso.org/sci/facilities/paranal/instruments/flames/inst/specs1.html} 
for an updated list and description of the GIRAFFE gratings currently used.}.

\begin{figure}
   \centering
   \includegraphics[width=\columnwidth]{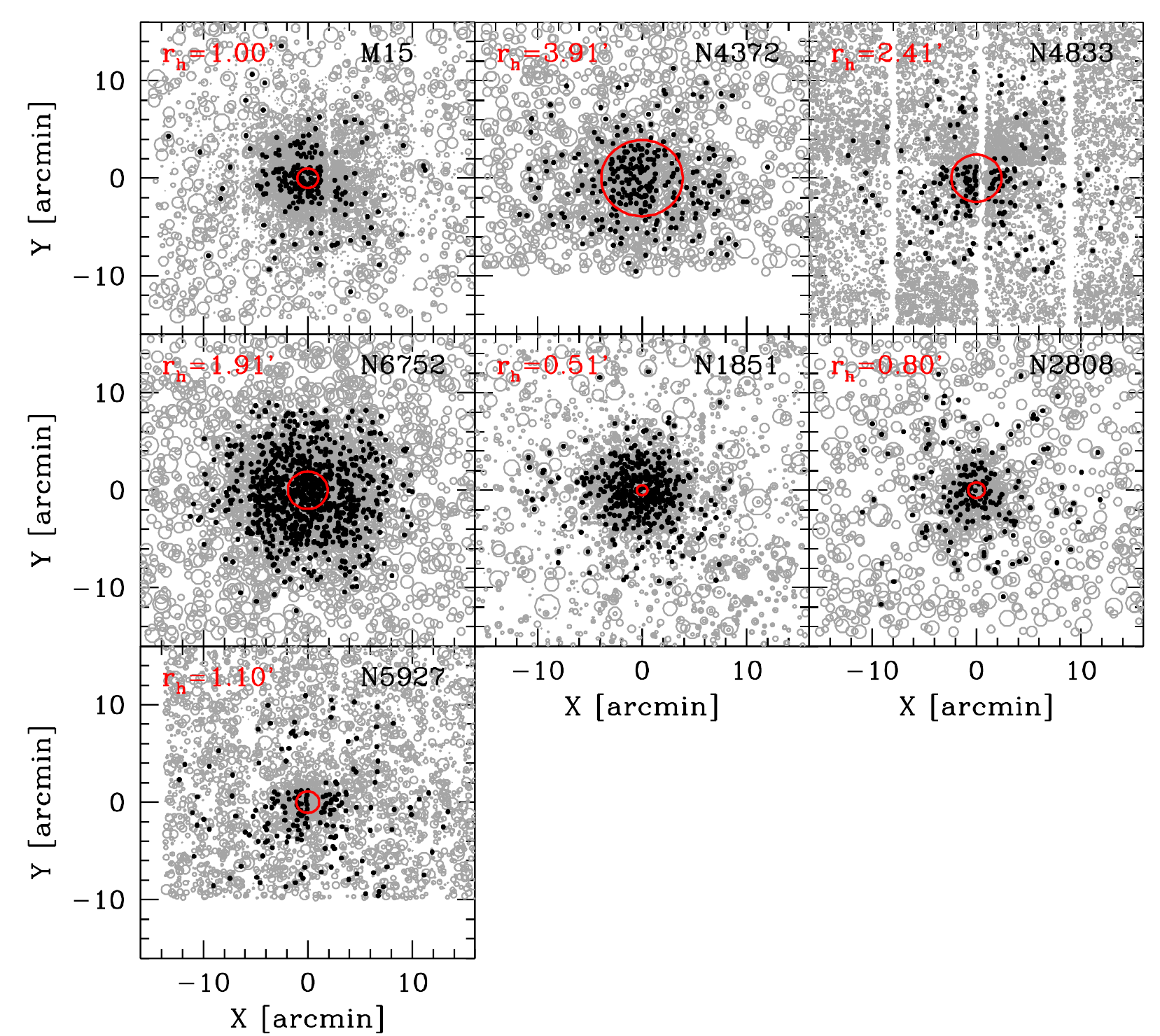}
   \caption{Spatial distribution of the initial sample (black dots)  overlaid on our WFI photometry (grey circles).
   The half-light radius (from ~\citealp{harris96}; 2010 edition) is also reported and plotted as a red line.
   Note that the stars plotted here are all the stars retrieved from the CASU archive before Galactic contaminants 
   were removed and sample selection was made (see text).}
              \label{XY}%
    \end{figure}

\begin{figure}
   \centering
   \includegraphics[width=\columnwidth]{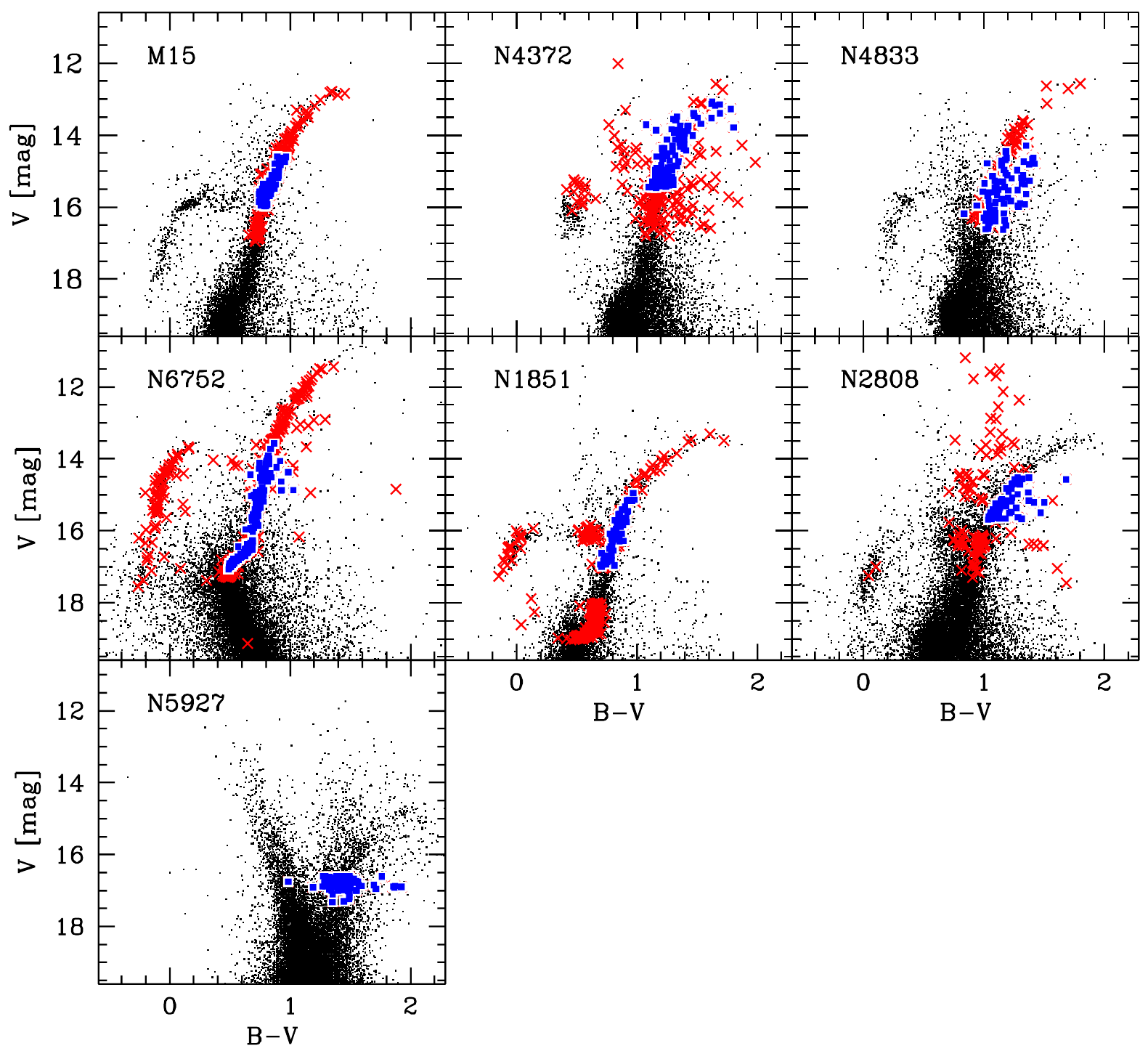}
   \caption{Gaia-ESO survey targets (blue squares) and GIRAFFE/FLAMES archival data (red crosses) overplotted on our WFI photometry (black dots).
    }
              \label{CMD}%
    \end{figure}

The GES and archival  spectra were processed by the survey pipeline (see Lewis et al., in preparation) 
and stored at the Cambridge Astronomical Survey Unit (CASU) Gaia-ESO Survey Archive (see Table~\ref{ARCHIVE} 
for a summary).  We present in Figs.~\ref{XY} and~\ref{CMD} the spatial distribution and the location on the cluster
colour-magnitude diagrams of the sampled stars.

While expanding our initial dataset, this exercise also allows us
to validate the results delivered by the survey data reduction pipeline.
We have for the entire sample at least 
two independent V$_{\rm{r}}$ estimates from observations with different instrument settings 
that we can use to check 
the internal consistency and accuracy of the derived radial velocities.
While complementing our data with archive data, we limited ourselves to samples that were
already incorporated by the Gaia-ESO survey pipeline when we started this analysis (February 2014). To maintain
the highest accuracy in the radial velocity estimates and the best homogeneity
in the velocity zero points, we included only samples of RGB stars that had stars
in common with the available sample of stars observed with the HR10 grating that is
the basis of our velocity scale (see below).

The data stored at the CASU Gaia-ESO Survey Archive
 are in multi-extension FITS files that contain both images with spectral 
data and tables with meta-data and derived information about each object, including 
radial heliocentric line-of-sight velocity measurements we used throughout this paper. 
In particular, radial velocities are measured using a two-steps approach.

The V$_{\rm{r}}$  determination is based on a procedure described in
\citet{koposov11}.
It uses direct per-pixel $\chi ^{2}$ fitting of
the spectra by templates. The main ingredient of the procedure is the generation of the
model spectrum, given log g, T$_{\rm{eff}}$, [Fe/H], and rotational velocity of
the star V$_{\rm{rot}}$.  For this purpose we used the template grid computed at
high resolution by \citet{munari05}.
The initial step of the V$_{\rm{r}}$ determination is the cross-correlation with the
subset of templates.  This step is only required to obtain a better initial
guess of the V$_{\rm{r}}$ and template for subsequent fit. 
The next step consists of a process of iteratively improving the stellar
template and V$_{\rm{r}}$ by direct modelling.  The process of improving the template involves keeping
the radial velocity fixed while performing the downhill Simplex \citep{NelderMead65}
optimisation of $\chi^2$ by improving stellar parameter
estimates: log g , T$_{\rm{eff}}$, [Fe/H], and V$_{\rm{rot}}$.  After this process has
converged, we perform the V$_{\rm{r}}$ optimisation by evaluating the template on a
grid of radial velocities and computing the $\chi ^{2}$ as a function of
radial velocity.  Then the stellar parameter step and RV steps are
repeated a few times until convergence. 
The calculation of the $\chi^2$ for each log g, T$_{\rm{eff}}$, [Fe/H],
V$_{\rm{rot}}$ and V$_{\rm{r}}$ also involves simultaneous continuum determination 
\citep{koposov11}, where the observed spectrum is assumed to be the
multiplication of the template and a fixed-degree polynomial of
the wavelength. As a result of the procedure, we derive $\chi^2$ as a function of V$_{\rm{r}}$ 
for the best-fit template, from which the pipeline determines
the V$_{\rm{r}}$ estimate and its uncertainty (see also \citealt{jeffries14}).

\begin{figure}
   \centering
   \includegraphics[width=\columnwidth]{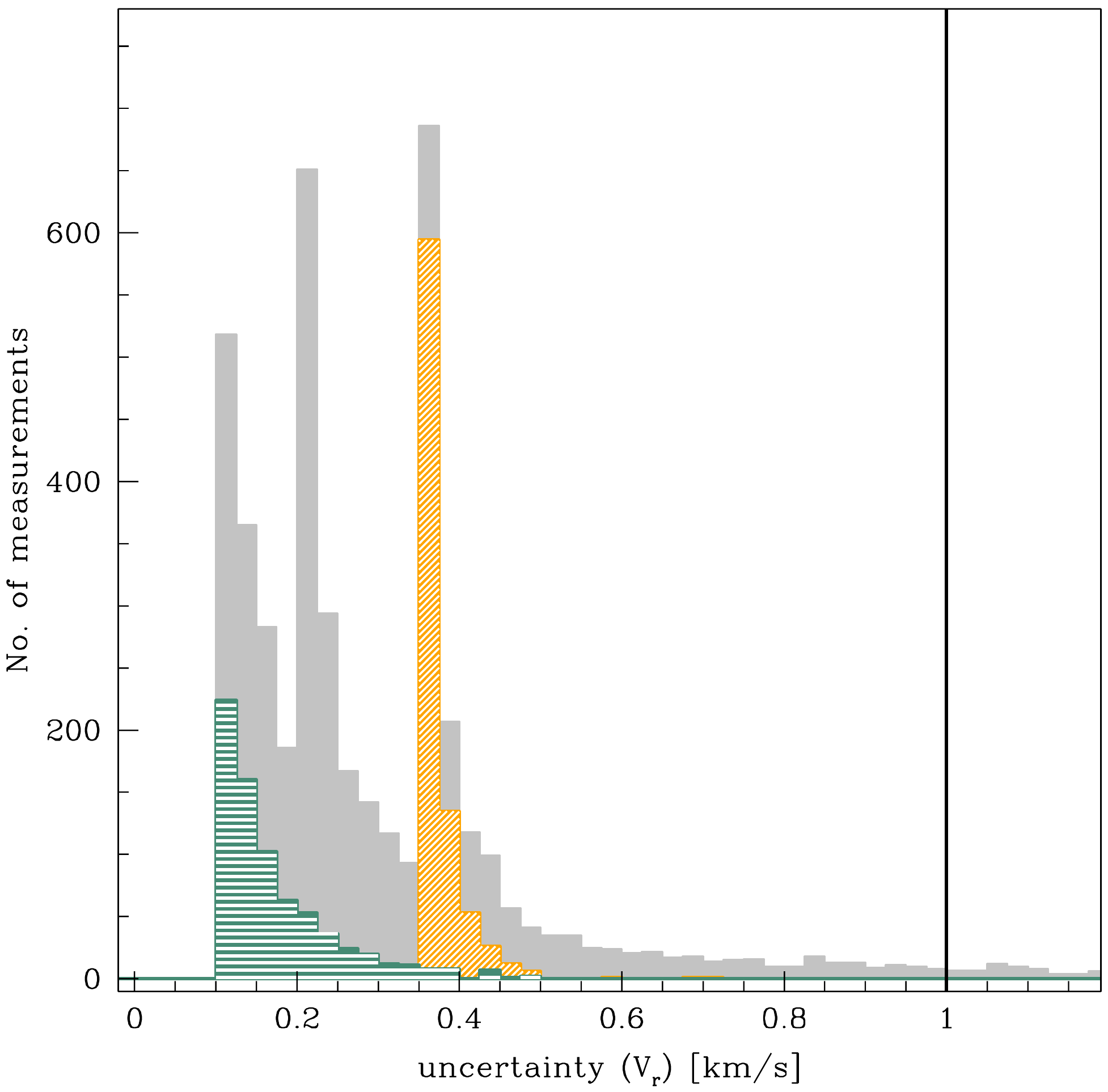}
   \caption{Distribution of velocity pipeline internal uncertainties associated with each V$_{\rm{r}}$ measurement, grey histogram, for all the considered
    settings, of HR 10, green histogram shaded at 0 degrees, and for HR 21, yellow histogram, shaded at 45 degrees. The vertical line is the adopted threshold for rejecting objects (see text).}
              \label{ERR}%
    \end{figure}

\subsection{V$_{r}$ estimates from repeated measurements}\label{RV}
As there are several stars in common between the observational datasets 
with different setups, we can check the internal consistency of the radial 
velocities delivered by the survey pipeline.
The mean (median) uncertainty value on individual pipeline V$_{\rm{r}}$ estimates is of 
0.17 (0.15)  and 0.38 (0.37)  km  s$^{-1}$ ($rms$ = 0.07 and 0.05, stars=731 and 830)
for the two Gaia-ESO setups HR 10 and HR 21, respectively (see Figure~\ref{ERR}).
The vast majority of the spectra ($\simeq$92\%) have uncertainties on V$_{\rm{r}}$ 
$\leq$ 1.0 km~s$^{-1}$, $\simeq$84\% $\leq$ 0.5 km~s$^{-1}$, 
small enough to not affect the measurement of the internal kinematics 
of the clusters. We decided to adopt a conservative threshold (uncertainty on V$_{\rm{r}}$ $\leq$ 1 km~s$^{-1}$)
to select the stars in the following analysis.

\begin{figure}
   \centering
   \includegraphics[width=\columnwidth]{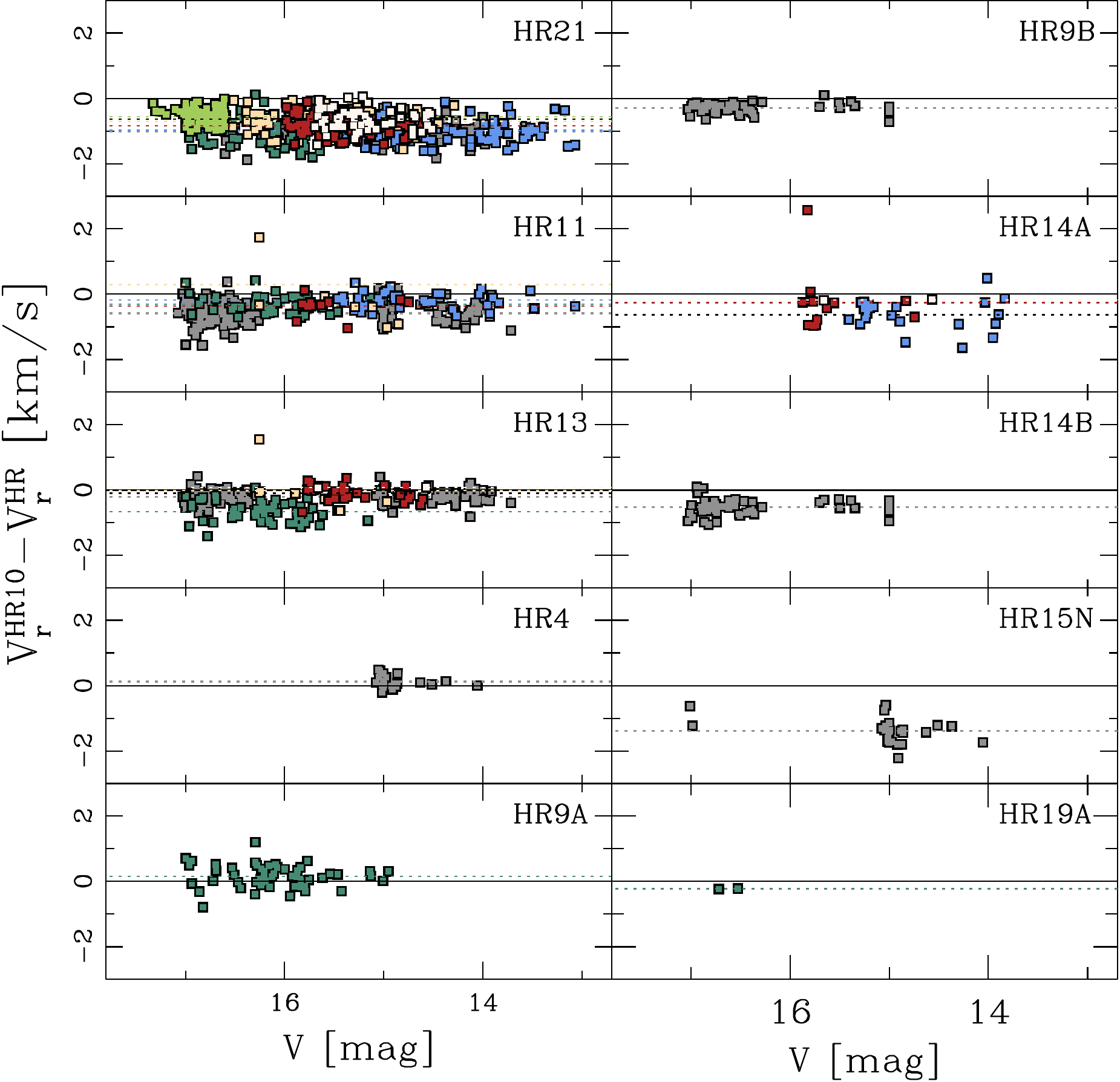}
   \caption{Comparison between the V$_{\rm{r}}$ estimated from spectra obtained with HR10 and 
   other GIRAFFE setups. Different colours correspond to different clusters: M15 (red), 
   NGC 4372 (light blue), NGC 4833 (apricot), NGC 6752 (grey), NGC 1851 (green), NGC 2808 (ivory),
and   NGC 5927 (light green). The dotted lines indicate the mean difference.}
              \label{GRAT}%
    \end{figure}

The comparison between the V$_{\rm{r}}$  estimates obtained from HR 10 and the other GIRAFFE setups 
for stars with uncertainty on  V$_{\rm{r}}$ $\leq$  1.0 km~s$^{-1}$ is shown in Fig.~\ref{GRAT} for all clusters.
Velocities from HR 10 were chosen as a reference because this setup is used,
together with HR 21, to observe all the stars targeted by Gaia-ESO survey, and their associated 
uncertainties are typically smaller than those of HR 21.
The mean difference and the standard deviation of the difference between the two sets of estimates 
with different setups are reported in Table~\ref{ZEROP}. The table also lists the number of 
stars in common between HR 10 and a given setup. 
Although the consistency among the different sets of measures 
is good (i.e., $\Delta$V$_{\rm{r}}$ $\leq$ 1.0 km~s$^{-1}$), we
note that there are differences in the V$_{\rm{r}}$ zero point (see also  
\citealp{donati14}). 
This might be due to the fact that Gaia-ESO survey HR 10 observations are generally
interleaved with a short exposure in which five dedicated fibres were illuminated by a bright 
(compared to the stellar spectra) thorium-argon (ThAr) lamp (see also \citealt{jeffries14}). 
These short exposures ({\it simcal} observations), combined with much longer day-time 
ThAr lamp exposures that illuminated all the instrument fibres, 
are used to adjust both the localisation and the wavelength solution, resulting in a higher precision in 
radial velocity determinations.
However, the differences in the zero-point between the 
ten V$_{\rm{r}}$ sets are not a reason for concern in the present analysis. 
In some cases, the comparison is based on only a handful of stars (see Fig.~\ref{ZEROP}), but 
because we did not detect trends and/or large spreads in the $\Delta$ V$_{\rm{r}}$, 
we decided to include these setups in the following analysis
as well.
The typical precision, as measured from the $rms$ of each set of $\Delta$ V$_{\rm{r}}$ 
computed after recursive clipping of the very few 3$\sigma$ outliers 
is $\leq$ 1.6 km~s$^{-1}$, but typically much lower than this, about 0.3 km~s$^{-1}$, 
which is more than satisfying for our purpose here. 
The actual uncertainty on the single measure should be smaller 
than the $rms$ of $\Delta$V$_{\rm{r}}$ because the latter includes the uncertainties 
of both estimates, added in quadrature. 

As a final step, we transformed all radial velocities into the HR 10 system by applying the shifts 
listed in Table~\ref{ZEROP} and weighting them by their uncertainty to derive the 
final V$_{\rm{r}}$. In the case of a single 
V$_{\rm{r}}$ determination we assigned the corrected V$_{\rm{r}}$ value to the star along with 
the formal uncertainty associated with the single measure.

As an additional validation of our final V$_{\rm{r}}$, we compared our determinations with those in the existing 
literature for NGC 6752, NGC 1851, and NGC 5927.
For NGC 6752, we found 159 stars in common with the sample presented by \citet{lane10b},
and for these stars we measured a mean difference 
V$_{\rm{r}}$ (this paper) - V$_{\rm{r}}$(Lane) of --0.95, $\sigma$=1.90 km~s$^{-1}$.
For NGC 1851 we have 104 stars in common with \citet{carretta1851}.
Our V$_{\rm{r}}$ determinations agree well with those from these authors ($\Delta$ V$_{\rm{r}}$ = 0.06, 
$\sigma$ = 0.7 km~s$^{-1}$). For NGC 5927 we measured a mean difference of 
V$_{\rm{r}}$(this paper)- V$_{\rm{r}}$(Simmerer)= --0.03,  $\sigma$=0.41 km~s$^{-1}$ for the stars in common
with the sample presented in \citet{simmerer13}.

\begin{table*}
\tiny
 \caption{The sample and its internal V$_{\rm{r}}$ accuracy}
 \renewcommand{\tabcolsep}{0.21cm}
 \label{ZEROP}
 \centering\begin{tabular}{l r r r r r }
            \hline\hline
                                 &                 &                                  &                                &                             &    \\     
Cluster          &   $\langle \Delta V_{\rm{r}}\rangle ^{\rm{HR10-HR21}} $ &  $\langle \Delta V_{\rm{r}}\rangle ^{\rm{HR10-HR11}} $  &   $\langle \Delta V_{\rm{r}}\rangle ^{\rm{HR10-HR13}} $ &   $\langle \Delta V_{\rm{r}}\rangle ^{\rm{HR10-HR4}} $ & $\langle \Delta V_{\rm{r}}\rangle ^{\rm{HR10-HR9A}} $\\          
                 &     ( km~s$^{-1}$)                                      &     ( km~s$^{-1}$)                                  &     ( km~s$^{-1}$)                                    &     ( km~s$^{-1}$)                                    &     ( km~s$^{-1}$)     \\
\hline
M 15      & --0.83 ($\sigma$ = 0.27, 78)  & --0.36 ($\sigma$ = 0.30, 12)   & --0.10 ($\sigma$ = 0.24, 26) &                                 &                             \\                 
NGC 4372  & --0.99 ($\sigma$ = 0.32, 100) &                                & --0.19 ($\sigma$ = 0.24, 42) &                                 &                             \\                 
NGC 4833  & --0.69 ($\sigma$ = 0.37, 77)  &   0.29 ($\sigma$ = 1.61, 8)    & --0.02 ($\sigma$ = 0.66, 8)  &                                 &                             \\                
NGC 6752  & --1.00 ($\sigma$ = 0.33, 108) & --0.58 ($\sigma$ = 0.34, 148)  & --0.20 ($\sigma$ = 0.22, 105)& 0.118 ($\sigma$ =0.190, 23)   &                                 \\                 
NGC 1851  & --0.94 ($\sigma$ = 0.39, 91)  & --0.30 ($\sigma$ = 0.25, 51)   & --0.67 ($\sigma$ = 0.31, 56) &                               & 0.15 ($\sigma$ = 0.35, 52)    \\
NGC 2808  & --0.63 ($\sigma$ = 0.32, 58)  &                                &   0.07 ($\sigma$ = 0.02, 2)  &                            &                              \\             
NGC 5927  & --0.56 ($\sigma$ = 0.21, 108) &                                &                           &                               &                              \\             
\hline
\hline
                     &                 &                                  &                                &                             &    \\     
Cluster   &   $\langle \Delta V_{\rm{r}}\rangle ^{\rm{HR10-HR9B}} $ &  $\langle \Delta V_{\rm{r}}\rangle ^{\rm{HR10-HR14A}} $  &  $\langle \Delta V_{\rm{r}}\rangle ^{\rm{HR10-HR14B}} $ &  $\langle \Delta V_{\rm{r}}\rangle ^{\rm{HR10-HR15N}} $  &  $\langle \Delta V_{\rm{r}}\rangle ^{\rm{HR10-HR19A}} $\\
                 &     ( km~s$^{-1}$)                                      &     ( km~s$^{-1}$)                                  &     ( km~s$^{-1}$)                                    &     ( km~s$^{-1}$)                                    &     ( km~s$^{-1}$)     \\
\hline
M 15      &                                   &--0.25 ($\sigma$ = 0.95, 12)      &                                      &                                        & \\
NGC 4372  &                                   &--0.62 ($\sigma$ = 0.48, 22)      &                                      &                                        & \\
NGC 4833  &                                   &                                           &                                  &                                         & \\
NGC 6752  &--0.29 ($\sigma$ = 0.15, 51)    &                                        &--0.54 ($\sigma$ = 0.23, 54)  &--1.39 ($\sigma$ = 0.37, 25)           & \\
NGC 1851  &                                   &                                     &                                         &                                        &--0.232 ($\sigma$ = 0.005, 2)\\
NGC 2808  &                                   &--0.175 ($\sigma$ = 0.01, 2)       &                                       &                                       & \\
NGC 5927  &                                   &                                     &                                         &                                       & \\            
\hline         

\hline         
            \end{tabular}

\end{table*}

\begin{figure}
   \centering

   \includegraphics[width=\columnwidth]{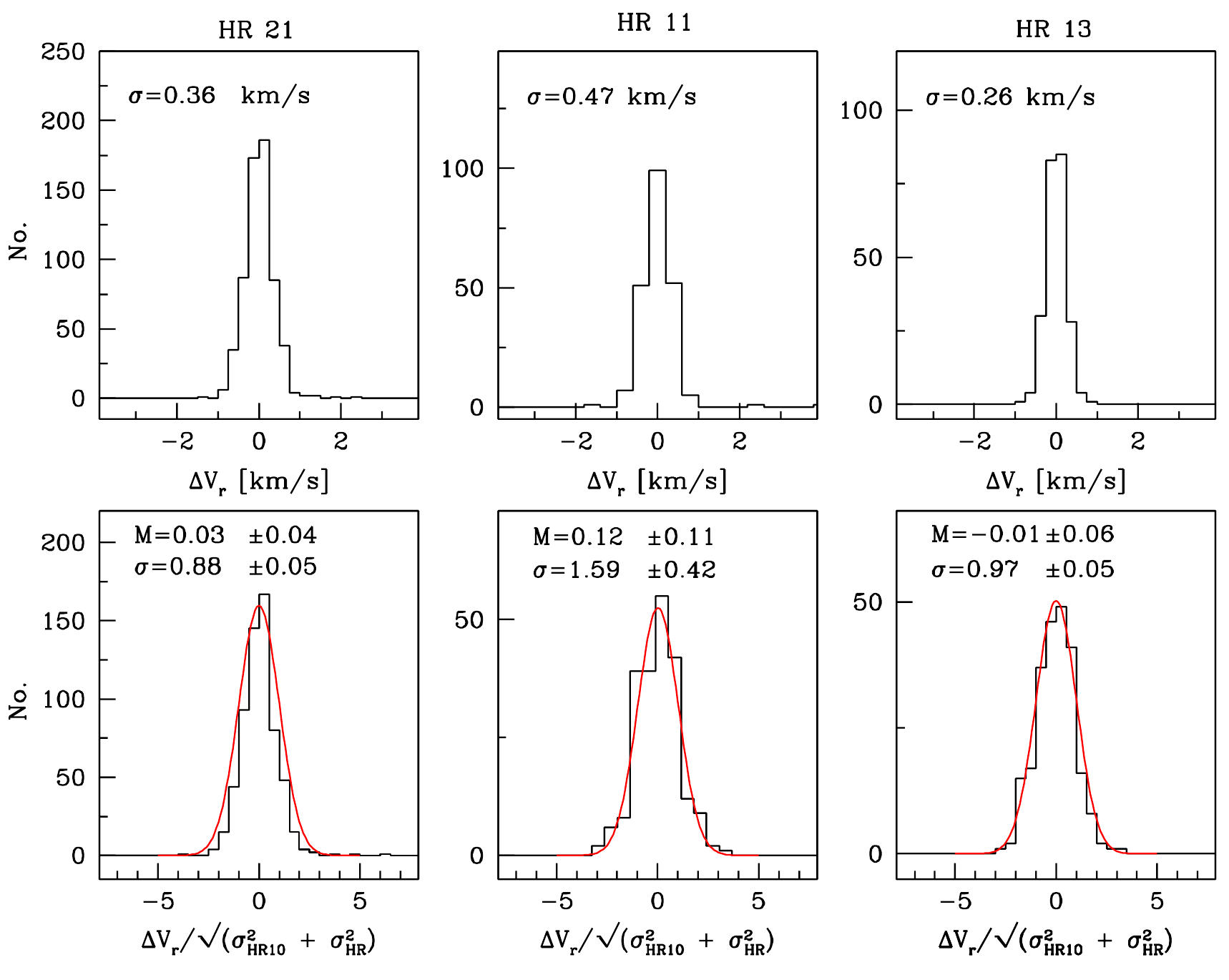}\\
   \caption{Comparison between velocity measurements for stars observed in two Giraffe setups.
  {\em Upper panels}: we show the distribution of velocity differences with respect to the velocity
  measured with HR 10 for all the stars observed (from left to right,  with HR 21, HR 11 and HR 13)
  and estimated uncertainties on V$_{\rm{r}}$ $\leq$ 1 km~s$^{-1}$ for each measurement.
  The mean difference and the $rms$ dispersion are also shown. {\em Bottom panels}: as above,  but now the velocity difference is normalised by the predicted uncertainty. It can be appreciated that 
  the measured uncertainty in the velocity distribution is very close to the unit variance Gaussian 
  (standard deviation = 0.88, 1.59, and 0.97 
  for HR 21, HR 11, and HR 13, respectively).}
              \label{GAUSS}%
    \end{figure}

    \begin{figure}
   \centering

   \includegraphics[width=\columnwidth]{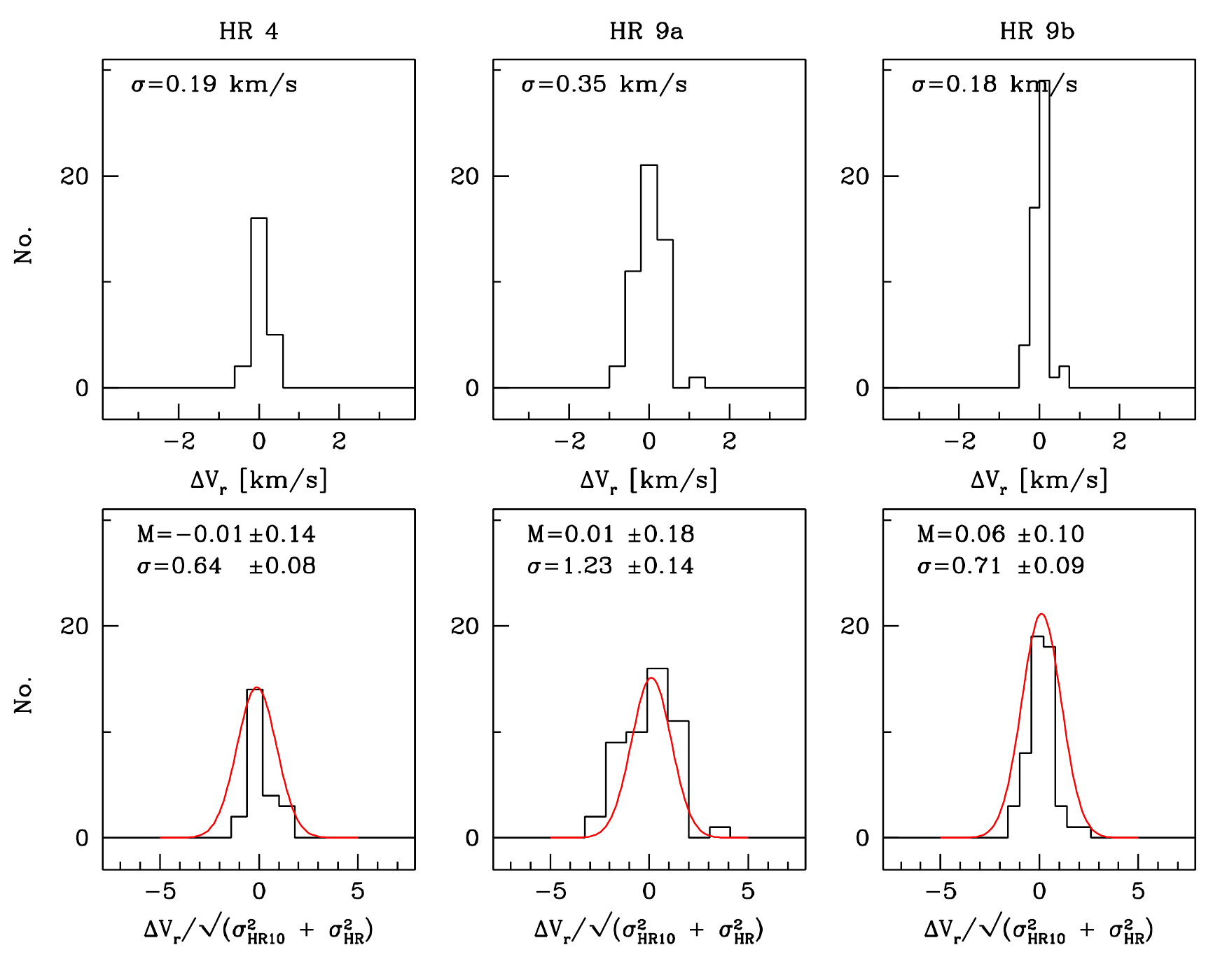}\\
   \caption{Same as Fig.~\ref{GAUSS}, but for HR 4, HR 9A, and HR 9B.}
              \label{GAUSS2}%
    \end{figure}

    \begin{figure}
   \centering

   \includegraphics[width=\columnwidth]{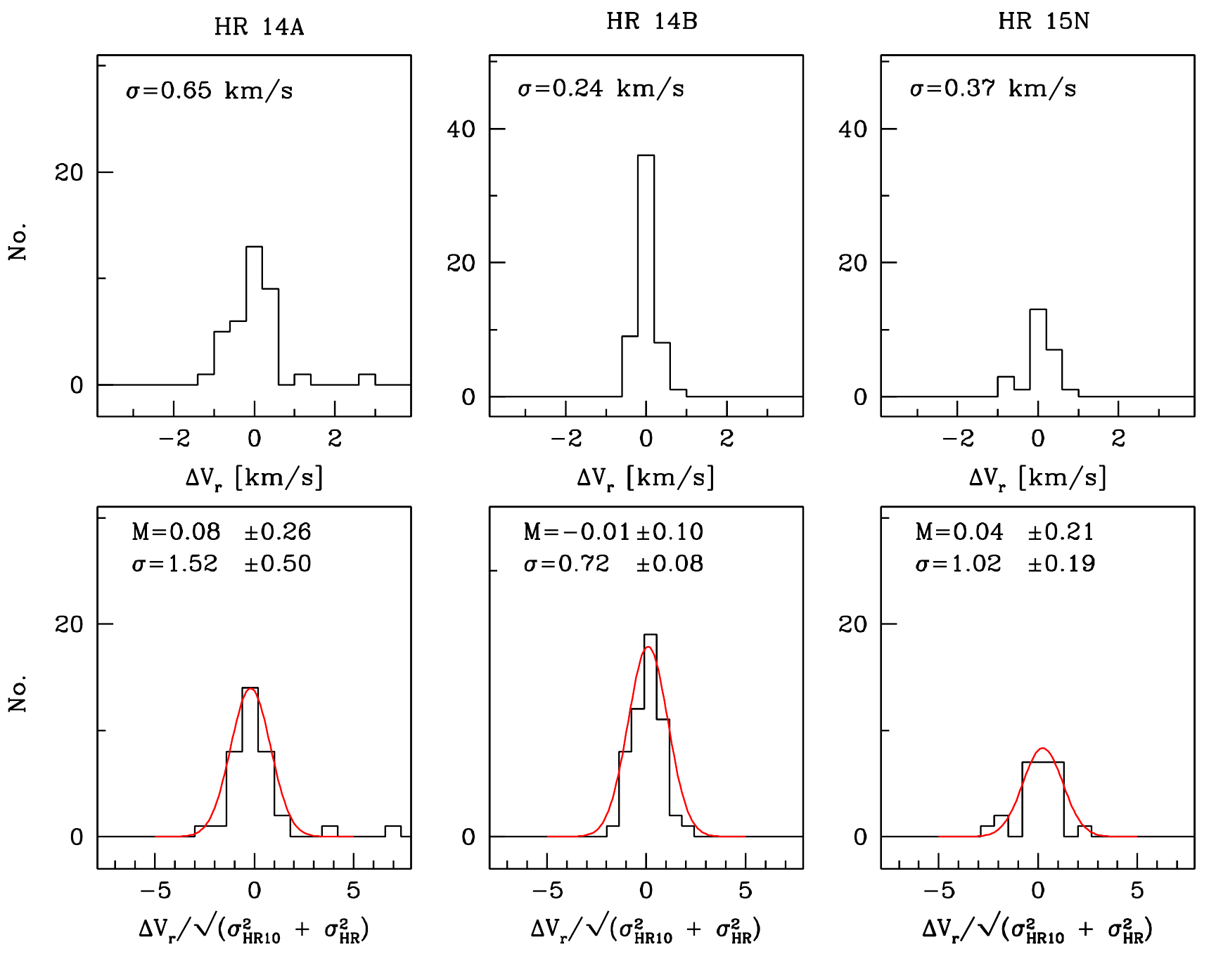}\\
   \caption{Same as Fig.~\ref{GAUSS}, but for HR 14A, HR 14B,
and HR 15N.}
              \label{GAUSS3}%
    \end{figure}    
    
\subsection{ V$_{r}$ errors from repeated measurements}\label{ERRORS}  
    
We tested the reliability of the pipeline-delivered V$_{\rm{r}}$ and their 
associated uncertainties by analysing the distribution of velocity differences from repeated measurements. 
We assumed that the {\em n}th observed velocity  $v_{n}$ (i.e., in different GIRAFFE setups)
can be considered a random variable that follows a Gaussian distribution centred on the 
true velocity value V$_{\rm{r}}$ and with dispersion given by the velocity uncertainty $\sigma _{n}$.
The difference between two repeated, independent measurements $v_{1}$ and $v_{2}$ , 
$\Delta v$ = $v_{1}$ -- $v_{2}$, is a random variable following a Gaussian distribution 
centred on zero and with a dispersion given by $\sigma$ = $\sqrt{\sigma^{2}_{1} + \sigma^{2}_{2}}$.
If both velocity and the related uncertainties are well determined, the distribution of velocity 
differences $\Delta v$ normalised by $\sigma$ should be a Gaussian with 
mean zero and dispersion unity.  
We considered all stars observed at least in two setups (i.e., HR 10 and another GIRAFFE setup) and
plotted the velocity differences and the normalised velocity differences for all the 
considered clusters\footnote{Each velocity estimate was previously 
corrected for the shifts listed in Table~\ref{ZEROP}.}.
Figure~\ref{GAUSS} shows that if we take into consideration all 
stars observed with HR 21, HR 11, and HR 13 (i.e., the setups for which we have the largest number 
of spectra available), all clusters have distributions with Gaussian appearance and dispersion 
equal to (or lower than) unity. 
We found that normalised $\Delta v$ distributions are all close to Gaussian, with a resulting 
standard deviation always smaller than 1.6, but typically equal
to or lower than unity for the remaining 
setups\footnote{We do not plot the comparison between velocity measurements for stars observed in both HR 10 
and HR 19A because there are only two stars in common between these two setups.}
(see Figs.~\ref{GAUSS},~\ref{GAUSS2}, and~\ref{GAUSS3}). We found 
higher $\sigma$ values for the setups that are commonly used for hot or rotating horizontal branch stars.

\subsection{Membership}\label{Member}

\begin{figure}
   \centering
   \includegraphics[width=\columnwidth]{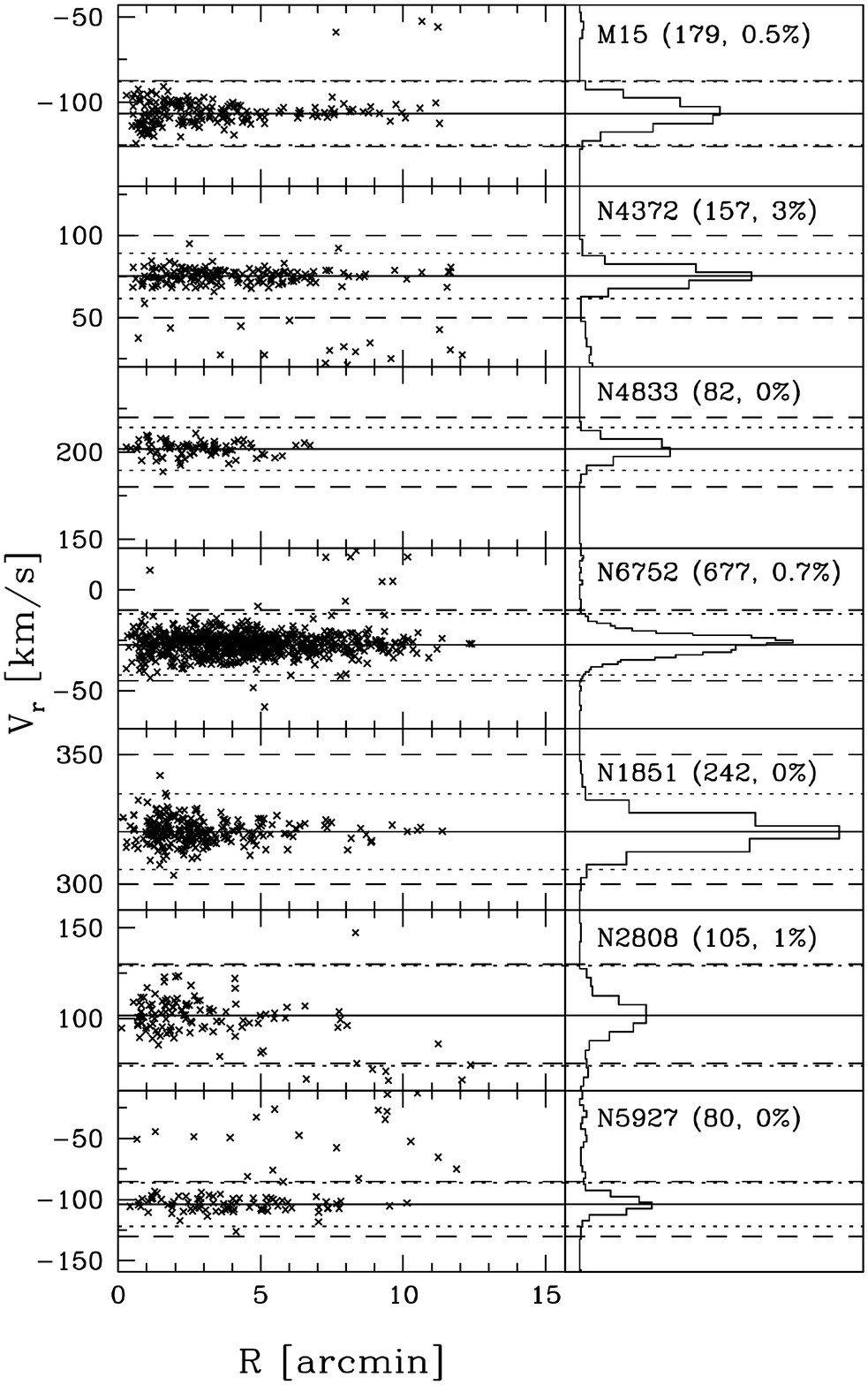}
   \caption{Radial velocity of program stars as a function of distance from the center ({\em left-hand panel}) for all the 
   considered clusters, and radial velocity distribution ({\em right-hand panel}). 
   The long-dashed lines mark the range we adopted for the first selection of candidate cluster members. 
   The dotted lines enclose the (global) $\pm$3$\sigma$ range from the mean of 
   the selected samples of candidates (continuous line), their number size is also indicated in the right-hand panel,
   along with the percentage of expected contaminants from the Besan\c{c}on models (see text).}
              \label{MEMBERSHIP}%
    \end{figure}

The distribution of the radial velocity of all the observed stars as a function of their 
(projected) distance from the centre
is shown in Fig.~\ref{MEMBERSHIP}. 
The coordinates of the cluster centre are taken from \citet{sh86} for NGC 4372 and NGC 4833,  \citet{go10} for NGC 5927, and 
\citet{noyola06} for the remaining clusters.
The distribution of radial velocities for the cluster 
members can be easily isolated from field contaminants in almost all cases.
Therefore, as a first broad selection, we kept as cluster members all stars 
with V$_{\rm{r}}$ between the two dashed lines in Fig.~\ref{MEMBERSHIP}.
We then computed the mean and dispersion of this sample and retained all stars with 
V$_{\rm{r}}$ within $\pm$3$\sigma$ range around the global 
mean (i.e., stars enclosed within the two dotted lines in the same figure).

 \citet{kouwenhoven08} demonstrated that even a binary fraction as high as 100 percent could 
lead to an increase in the observed velocity dispersion to lower than $\leq$ 0.5 km~s$^{-1}$.
Since GCs have typical binary fractions $\leq$ 20 percent (i.e., \citealp{sollima07} and \citealp{milone08}),
we considered binaries as a negligible factor for our analysis.
We expect some (limited) contamination from Milky Way stars, even in our V$_{\rm{r}}$ 
-selected sample. We used the Besan\c{c}on model \citep{robin03} to 
simulate a set of V$_{\rm{r}}$  for stars that correspond to the direction, colour, and magnitude survey of the targets.
The Besan\c{c}on model suggests that some spurious Milky Way contaminant 
may be present even in the relatively narrow V$_{\rm{r}}$ range we have adopted to select stars. 
In the right-hand panel of Fig.~\ref{MEMBERSHIP}, we 
show the histograms of the distribution of the  V$_{\rm{r}}$ for each cluster, the number of stars
selected as possible cluster members, and the (small percent) contamination expected according to \citet{robin03}
Galactic model. Finally, in the following sections, we reconsider individual memberships based on the velocity
distributions as a function of distance from the cluster centre.

\section{Velocity dispersion profiles}\label{sigma}

Although all clusters we studied  have kinematic data already available in the literature (for an update summary we refer to 
Table~1 of B12),  there are a few clusters for which we can provide a significant improvement over existing kinematic data and analyses. 
For example, while M 15  has been extensively studied (\citealp{vandenbosch06} presented a detailed analysis of this cluster based on nearly  two thousand 
V$_{\rm{r}}$ and proper motions), for NGC~5927 no velocity dispersion profile and no estimate of the central velocity dispersion are available in the literature 
(\citealp{simmerer13} provided only an estimate of the overall dispersion). For several clusters the samples presented in the literature are smaller than
(NGC 6752, NGC 1851;  \citealp{lane10b, scarpa11,carretta1851,car11})
or similar to (NGC 4833, NGC 4372; \citealp{carretta14,kacharov14}) those considered here. An independent check of the results from previous analyses is provided.
In the following we briefly discuss the properties of the V$_{\rm{r}}$ distributions and derive new estimates 
of the central velocity dispersion ($\sigma _{0}$)  in all the selected clusters.

\begin{figure}

   \centering
   \includegraphics[width=\columnwidth]{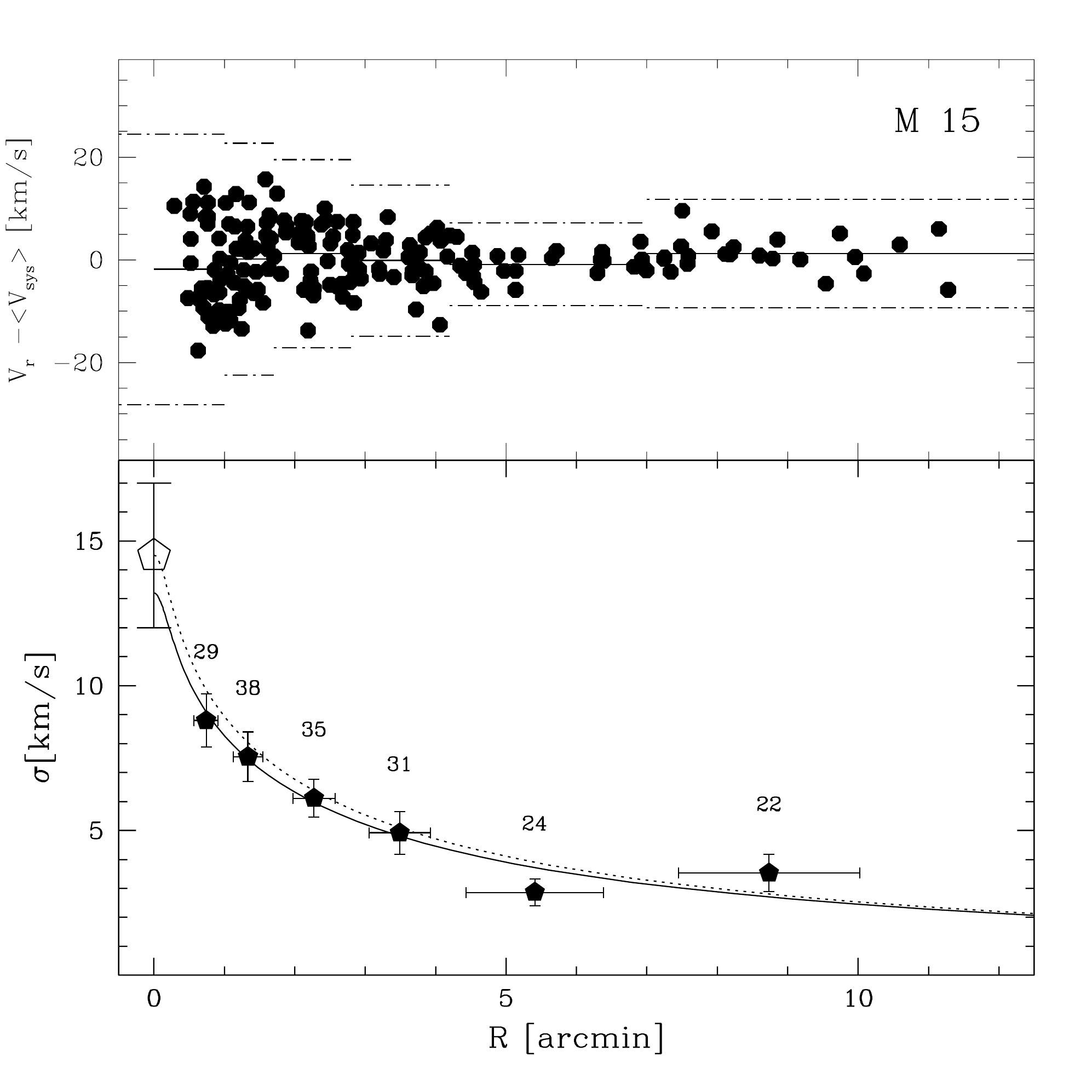}
   \caption{Velocity dispersion profile of M 15 stars. 
   The upper panel shows the V$_{\rm{r}}$ distribution as a function of distance from the cluster 
   centre for individual stars of the sample. Only stars plotted as dots are 
   retained to compute $\sigma$ in the various radial bins: 
   crosses are stars rejected only because they are {\em local} 3$\sigma$ 
   outliers of the bins. The mean V$_{\rm{r}}$ - $\langle$ V$_{\rm{sys}}$ $\rangle$ is marked by the continuous horizontal line. 
   Comparison of the observed velocity dispersion profile of M15 with the King model with a core radius r$_{C}$= 0.07\arcmin and  a concentration C=2.5, from \citet{trager93} and normalised 
   to  $\sigma _{0}$=13.2 km~s$^{-1}$ (continuous line; our estimate) and  $\sigma _{0}$=14.5 km~s$^{-1}$ (dotted line; by \citealp {mcnamara03}).
   The large filled pentagons are the dispersions estimated in the corresponding bins displayed in 
   the upper panel, with their bootstrapped errors. 
   The number of stars per bin is also reported above the points. 
   The open pentagon is the value of $\sigma$ at the centre of M 15 from \citet{mcnamara03}.}
   \label{m15}%
    \end{figure}

\begin{figure}[!h]
   \centering
   \includegraphics[width=\columnwidth]{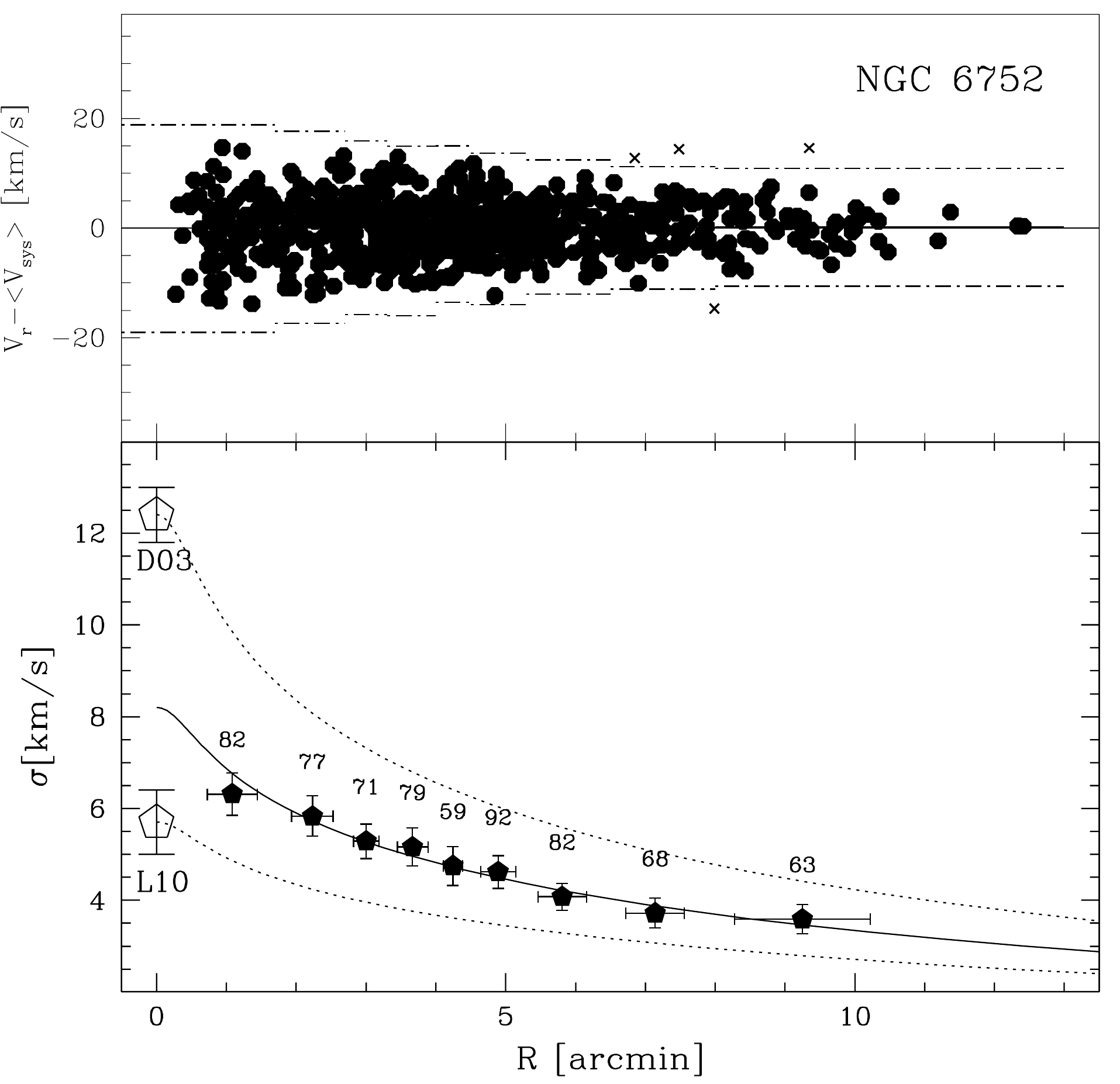}
   \caption{Same as in Fig.~\ref{m15}, but for NGC 6752. 
    Core radius  (r$_{C}$= 0.17\arcmin) and concentration (C=2.5) are from \citet{trager93} and K66 models are normalised 
   to  $\sigma _{0}$=8.2 km~s$^{-1}$ (continuous line; our estimate) and  
   $\sigma _{0}$=5.7 and 12.4 km~s$^{-1}$ (dotted lines; by \citealp {lane10b}~(L10) and \citealp{drukier03}~(D03)).
   The large open pentagons are the values of $\sigma$ at the centre from 
   L10  and D03. }
   \label{n6752}%
    \end{figure}

\begin{figure}[!h]
   \centering
   \includegraphics[width=\columnwidth]{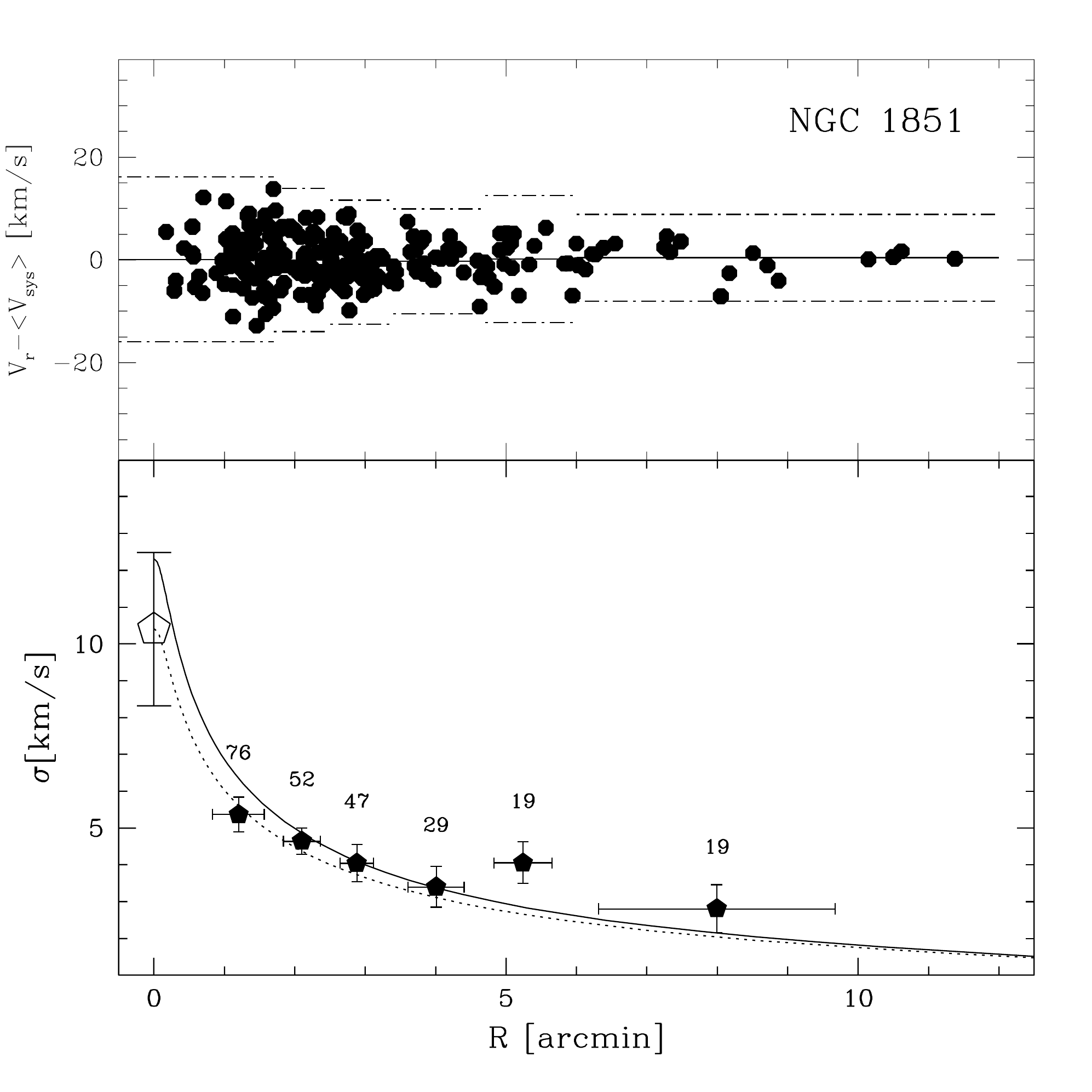}
   \caption{Same as in Fig.~\ref{m15}, but for NGC 1851.
   Core radius  (r$_{C}$=0.08\arcmin) and concentration (C=2.24) are from \citet{trager93} and K66 models are normalised 
   to  $\sigma _{0}$=12.3 km~s$^{-1}$ (continuous line; our estimate) and  
   $\sigma _{0}$=10.4 km~s$^{-1}$ (dotted line; by \citealp{pryor93}).
   The open pentagon is the value of $\sigma$ at the centre from \citet{pryor93}. }
   \label{n1851}%
    \end{figure}

\begin{figure}[!h]
   \centering
   \includegraphics[width=\columnwidth]{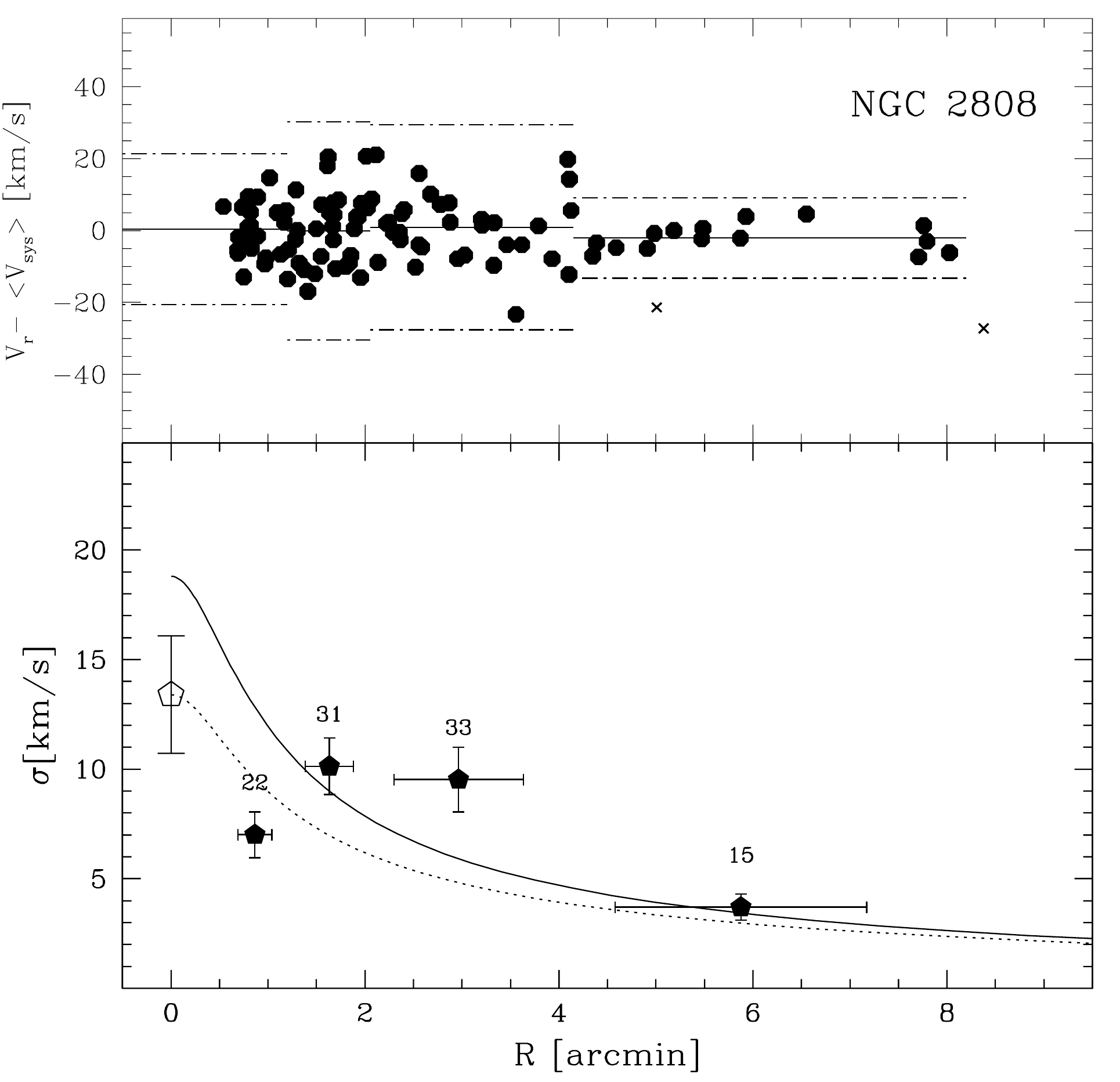}
   \caption{Same as in Fig.~\ref{m15}, but for NGC 2808.
   Core radius  (r$_{C}$=0.26\arcmin) and concentration (C=1.8) are from \citet{trager93} and K66 models are normalised 
   to  $\sigma _{0}$=18.8 km~s$^{-1}$ (continuous line; our estimate) and  
   $\sigma _{0}$=13.4 km~s$^{-1}$ (dotted line; by \citealp{pryor93}).
   The open pentagon is the value of $\sigma$ at the centre from \citet{pryor93}. }
   \label{n2808}%
    \end{figure}

\begin{figure}[!h]
   \centering
   \includegraphics[width=\columnwidth]{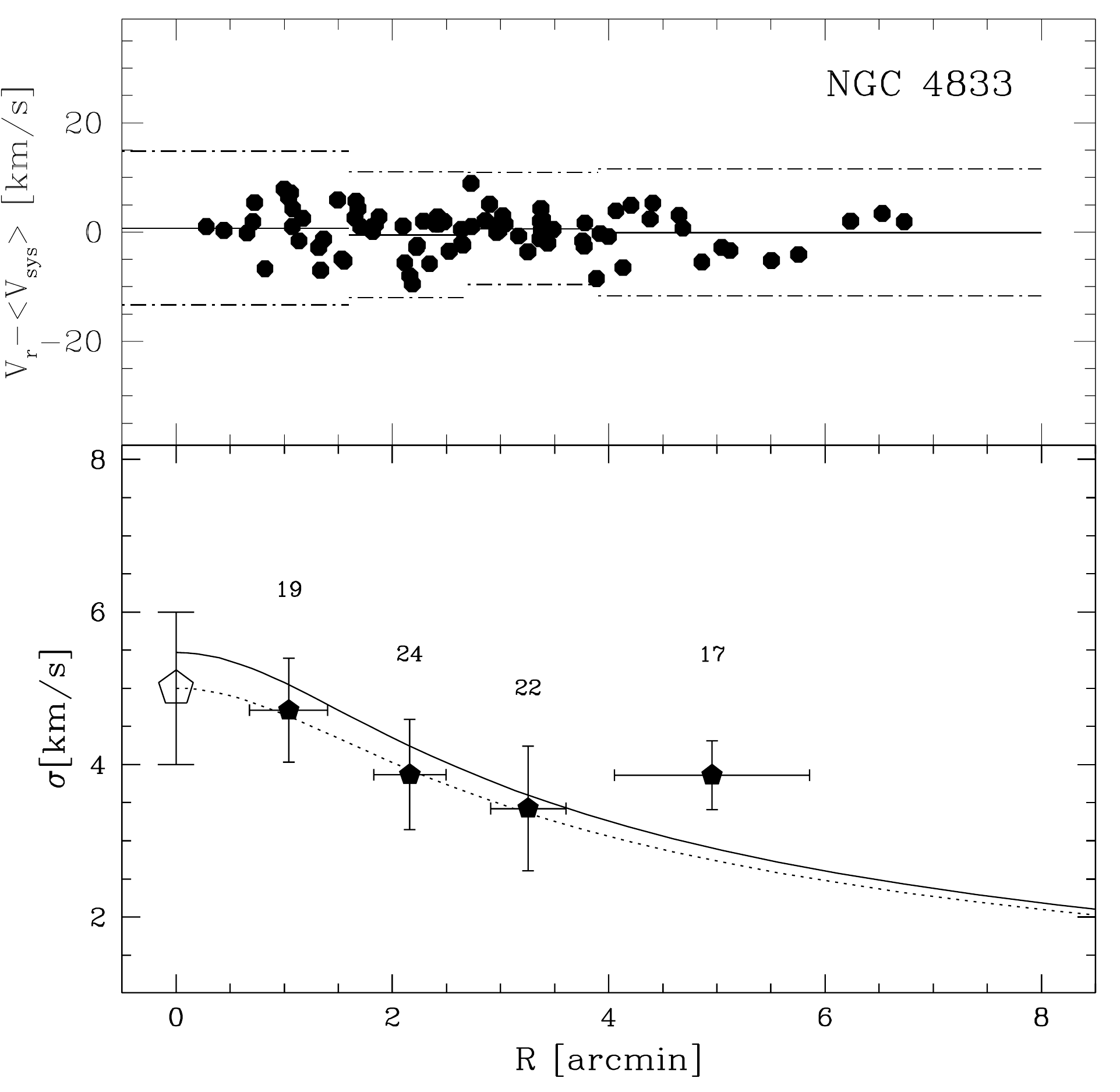}
   \caption{Same as in Fig.~\ref{m15}, but for NGC 4833. 
   Core radius  (r$_{C}$=1.0\arcmin) and concentration (C=1.25) are from \citet{trager93} and K66 models are normalised 
   to  $\sigma _{0}$=5.5 km~s$^{-1}$ (continuous line; our estimate) and  
   $\sigma _{0}$=5.0 km~s$^{-1}$ (dotted line; by \citealp{carretta14}).
   The open pentagon is the value of $\sigma$ at the centre from \citet{carretta14}. }
   
   \label{n4833}%
    \end{figure}

\begin{figure}[!h]
   \centering
   \includegraphics[width=\columnwidth]{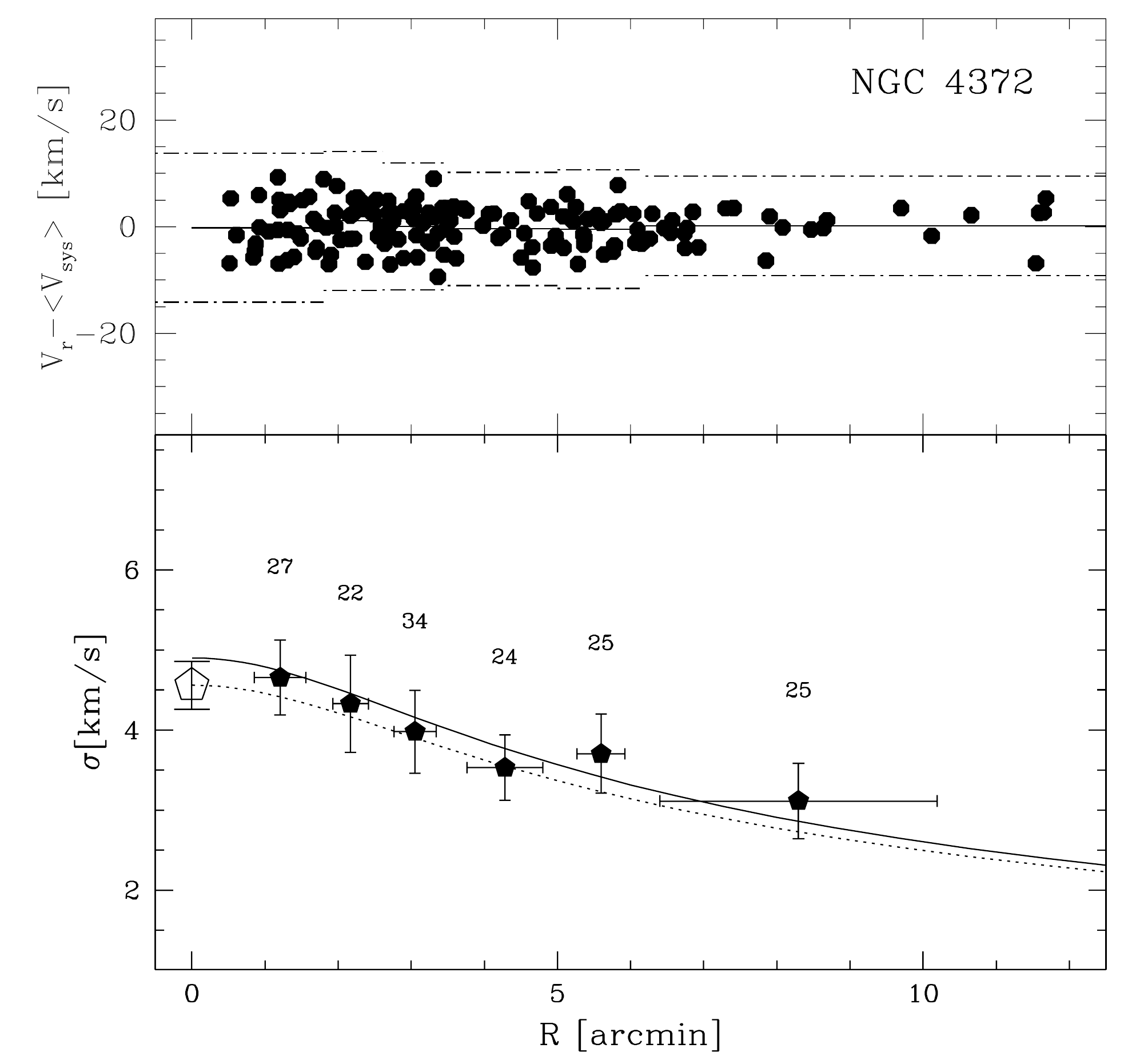}
   \caption{Same as in Fig.~\ref{m15}, but for NGC 4372. 
    Core radius  (r$_{C}$=1.74\arcmin) and concentration (C=1.30) are from \citet{trager93} and K66 models are normalised 
    to  $\sigma _{0}$=4.9 km~s$^{-1}$ (continuous line; our estimate) and  
    $\sigma _{0}$=4.56 km~s$^{-1}$ (dotted line; by \citealp{kacharov14}).
   The open pentagon is the estimate of $\sigma$ at the centre 
   from \citet{kacharov14} based on the fit of a Plummer profile and a rotating, physical model.}
   \label{n4372}%
    \end{figure}

\begin{figure}[!h]
   \centering
   \includegraphics[width=\columnwidth]{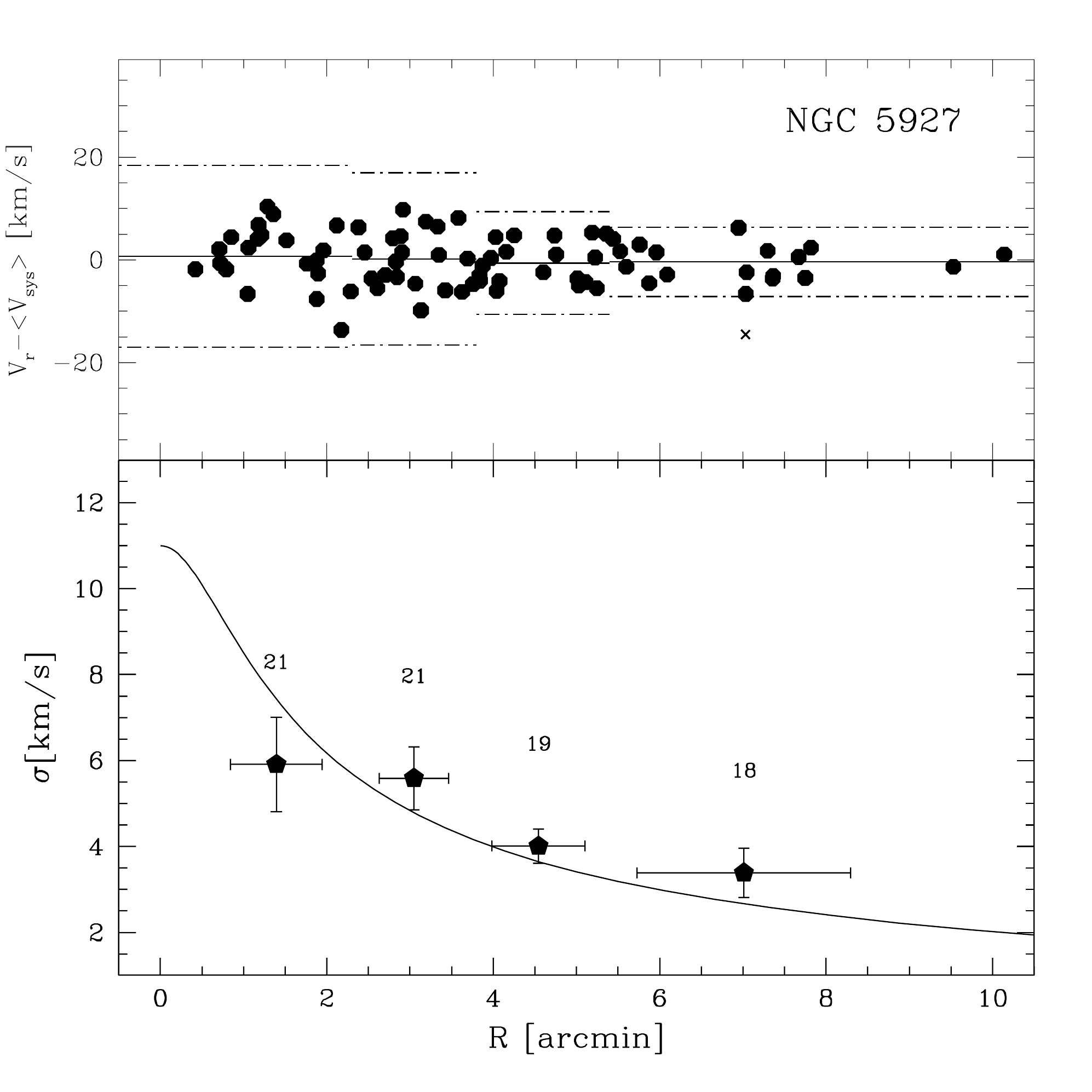}
   \caption{Same as in Fig.~\ref{m15}, but for NGC 5927.
   Core radius  (r$_{C}$=1.40\arcmin) and concentration (C=1.60) are from \citet{trager93} and K66 models are normalised 
    to  $\sigma _{0}$=11.0 km~s$^{-1}$ (continuous line; our estimate).}
   \label{n5927}%
    \end{figure}

We used radial velocities of member stars to produce velocity dispersion ($\sigma$) curves 
for all the considered clusters as described in \citet{bellazzini08}, 
using {\em jackknife} resampling \citep{lupton93} to compute uncertainties. 
In the upper panel of Figs.~\ref{m15} to~\ref{n5927} 
we show the V$_{\rm{r}}$ distribution as a function of R (distance from the centre). 
We divided the whole sample into several independent radial bins of different size,
manually chosen as a compromise between maintaining the highest degree of spatial resolution 
while considering a statistically significant number ($\simeq$ 15) of stars.
In each bin we computed the average V$_{\rm{r}}$ - $\langle$ V$_{\rm{sys}}$ $\rangle$ and velocity dispersion $\sigma$, 
with their uncertainties. An iterative 3$\sigma$ clipping algorithm was applied 
bin by bin. Any star rejected by the clipping algorithm 
was then rejected from the following analysis. The rejected stars are indicated in the plots as crosses.
The V$_{\rm{r}}$ estimates for all the stars judged to be members are reported in Table~\ref{VELOCITA},
together with other stellar parameters. 
In Table~\ref{VELOCITY} we report the measured average velocity for each cluster.
From this table we note an excellent agreement  between the cluster average velocity derived here and those reported in literature.

\begin{table}
\caption{Radial velocities for the stars.} 
\renewcommand{\tabcolsep}{0.13cm}

\label{VELOCITA}
\centering 
\begin{tabular}{c c c c c c c }
\hline\hline 
NGC    & ID  &  RA    & Dec   & V      & V$_{\rm{r}}$ & eV$_{\rm{r}}$ \\
       &     &  (deg) & (deg) & (mag)  & km~s$^{-1}$ & km~s$^{-1}$ \\
       \hline
7078   &  1  & 322.4817397  & 12.1793098 & 12.8 & -118.90 & 0.64 \\
7078   &  2  & 322.5093355  & 12.1893088 & 12.8 &  -98.25 & 0.24  \\
7078   &  3  & 322.5037366  & 12.1491900 & 12.9 & -114.20 & 0.38 \\
7078   &  4  & 322.5013943  & 12.1808019 & 13.0 & -116.60 & 0.41  \\
7078   &  5  & 322.4908124  & 12.1577422 & 13.2 &  -95.12 & 0.24 \\ 
7078   &  6  & 322.4993224  & 12.1571307 & 13.3 & -112.00 & 0.11 \\ 
   \hline 
\end{tabular}
\tablefoot{A portion of the table is shown for guidance about its content, the complete table is available 
in electronic format through the CDS service.}

\end{table}

 The derived velocity dispersion profile is reported in the lower panel of the figures and listed in Table~\ref{AMMASSI}. The profiles are complemented with the central estimate obtained from the literature (large empty pentagon in the same figures). 
 
 We fitted the resulting velocity dispersion profile in a least-squares 
 sense with the predictions of the \citet{king66} (hereafter K66) model that best fits the surface brightness profile \citep[according to][]{trager93}, leaving the central velocity dispersion $\sigma _{0} $as the free parameter to be determined.
 It is important to note that our $\sigma _{0}$ estimates are extrapolations to r=0 of the isotropic single-mass K66 model that best fits the observed velocity dispersion profile. 
 Hence they are model-dependent and based on models that are known not to be  perfectly adequate to describe real clusters, which, for instance, are populated by stars of different masses. 
 The reliability of each estimate of $\sigma _{0}$ depends on the radial coverage of the velocity dispersion profile and on the cluster surface brightness profile; it can be judged relatively easily from inspecting Figs.~7-13 below.

 \begin{table*}
\caption{Comparison between the systemic radial velocities derived in this paper with literature values.}  
\renewcommand{\tabcolsep}{0.25cm}

\label{VELOCITY}
\centering 
\begin{tabular}{l r r r r  l}
\hline\hline 
Target   &  V$_{\rm{r}}$ (t.p.) &  dispersion (t.p.)  &V$_{\rm{r}}$ (lit.)  &  dispersion (lit.)  & References \\  
             &    km~s$^{-1}$        &    km~s$^{-1}$    &    km~s$^{-1}$        &    km~s$^{-1}$ &  \\
\hline
M 15     &  --106.4$\pm$0.7    & 6.2   & --106.7$\pm$0.4   & 11.8   & \citet{mcnamara03}\\
NGC 1851 &    320.1$\pm$0.3    &4.4    &      320.3$\pm$0.4  & 3.7 & \citet{carretta1851}\\
NGC 2808 &    101.4$\pm$1.0    &9.5    &      102.4$\pm$0.9  & 9.8  & \citet{carretta06}\\
NGC 4372 &      75.2$\pm$0.4    & 3.9   &       75.9$\pm$0.4   & 3.8    & \citet{kacharov14}\\
NGC 4833 &     202.1$\pm$0.6   & 3.9   &     202.0$\pm$0.5   & 4.1    & \citet{carretta14}\\
NGC 5927 & --103.95$\pm$0.7  & 5.1   & --104.0$\pm$0.6     & 5.0  & \citet{simmerer13}\\
NGC 6752 &    --26.9$\pm$0.2   &5.0    &    --26.1$\pm$0.2     &4.7    & \citet{lane10b} \\

\hline 
\end{tabular}
\end{table*}

In general,  our $V_r^{sys}$ and the $\sigma _{0}$ estimates agree well with those found in previous studies (see Table~4), except for two cases.

For NGC 6752 we estimated a velocity dispersion toward the centre of $\sigma_{0}$ = 8.2 km~s$^{-1}$, which is higher 
than that found by \citet{lane10b}  ($\sigma_{0}$ = 5.7 $\pm$ 0.7 km~s$^{-1}$)\footnote{For reference
\citet{dubath97} obtained $\sigma_{0}$ = 4.9 $\pm$ 2.4 km~s$^{-1}$ from integrated-light spectra.}.
This can be partially due to the fact that they estimated $\sigma_{0}$ by extrapolating from a different
class of models than we did here, that is, \citet{plummer11} instead of K66. 
Our observed velocity dispersion profile is fully compatible with that by \citet{lane10b} in the wide
range where the two profiles overlaps. The inspection of the two curves suggests that the true value of $\sigma_0$ can be in between the two estimates.
On the other hand, the two estimates based on radial velocities are significantly lower than the one consistently derived from the two components of the proper motions in the plane of the sky by 
\citet{drukier03} ($\sigma_{0}$ = 12.4 $\pm$ 0.5 km~s$^{-1}$; see Fig.~\ref{n6752}).
This large discrepancy with the \citet{drukier03} {\em measured} value can be due to the adoption of a cluster distance that overestimates the true value, to a significantly different mean 
mass of the adopted tracers (e.g., giants vs. subgiants+dwarfs), or to a significant amount of orbital anisotropy (see \citealp{drukier03}). In any case, our data provide the final proof that the discrepancy between the dispersion from radial velocity and from proper motions, already noted by \citet{drukier03} is real and requires further investigation.

For NGC 2808, the sparse dispersion profile we obtained provides only weak constraints on $\sigma _{0}$, hence the difference between our extrapolated value and the value listed in \citet{pryor93} cannot be considered significant. We recall that the latter is from an integrated spectrum taken at the cluster centre, and it fully agrees with the recent measurement by \citet{lutz12}. 

For NGC 5927 we present for the first time a velocity dispersion profile in Fig.~\ref{n5927}. We also provide the first estimate of $\sigma_0$, but we note that the constraint on this 
parameter provided by our profile is relatively weak, hence the associated uncertainty is quite large (of about 2 km~s$^{-1}$).

 \section{Rotation}\label{rotazioni}

\begin{figure}
   \centering
   \includegraphics[width=6.8cm]{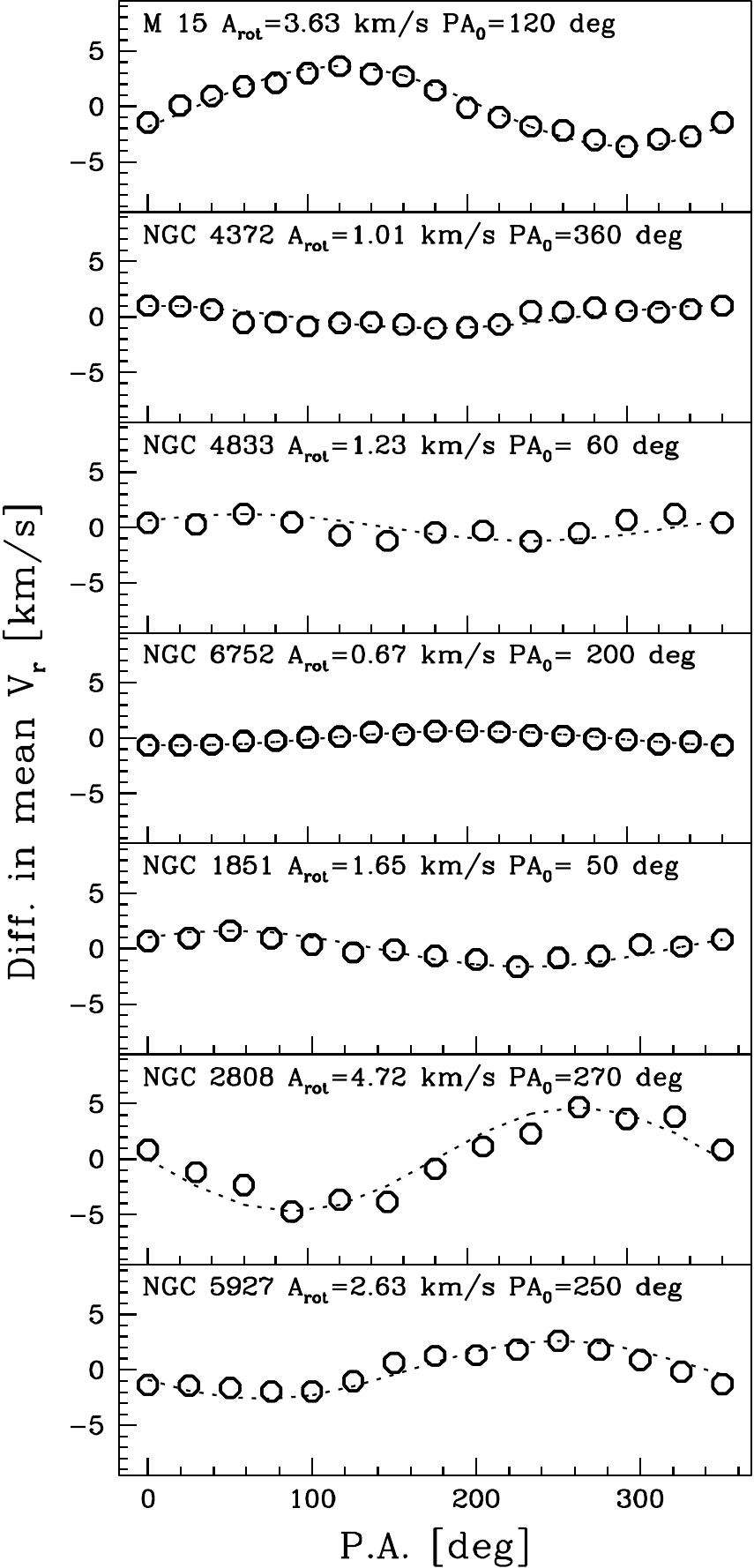}
   \caption{Rotation in our program GCs. The plots display the difference
   between the mean velocities of each side of a cluster with respect 
   to a line passing through the cluster centre with a varying PA (measured from north to east), as 
   a function of the adopted PA. The dashed line is the sine law that best fits the observed pattern.
   The rotational amplitude (A$_{\rm{rot}}$) and the position angle (PA) are also indicated.}
              \label{ROTAZIONI}%
    \end{figure}

We used our sample to search for a rotation signal in all the considered clusters.
To do this, we used the same method as adopted by \citet{cote95}, \citet{pancino07},
\citet{lane09,lane10a,lane10b}, and B12.
Rotations were measured by halving 
the cluster by position angle (PA)\footnote{In the adopted approach PA is defined to increase 
anti-clockwise in the plane of the sky 
from north (PA = 0$^{\circ}$) toward east (PA = 90$^{\circ}$).} and calculating the mean 
radial velocity of each half. 
This was performed in steps of 20-35$^{\circ}$ depending on the number of the observed stars in the 
considered cluster to avoid aliasing effects.
The two mean velocities were then subtracted, and the difference in the mean V$_{\rm{r}}$
for each PA of the dividing line is plotted in Fig.~\ref{ROTAZIONI} as a function of the PA and 
the best-fitting sine function 

$$\Delta \langle V_{\rm{r}} \rangle = A_{\rm{rot}} \sin (PA + \Phi),$$
where $\Phi = 270^{\circ}$ -- PA$_{0}$, PA$_{0}$ is the position angle of the dividing 
line corresponding to the maximum rotation amplitude (degrees), and A$_{\rm{rot}}$ is twice the actual 
mean amplitude (in km~s$^{-1}$; see \citealp{lane10a} and B12).
A$_{\rm{rot}}$/2 should be considered as an underestimate of the 
maximum projected rotational amplitude because the $\langle V_{\rm{r}} \rangle$ difference is actually averaged over 
the full range of radial distances covered by the targeted stars, and the amplitude does vary with distance 
from the cluster centre \citep{sollima09}.
But even if the derived A$_{\rm{rot}}$ are only estimates of 
the amplitude of the projected rotation pattern, 
we can consider A$_{\rm{rot}}$ as a proxy for the true amplitude, in a statistical sense (see Appendix A in B12). 
The estimates of A$_{\rm{rot}}$ should be considered as quite robust. 
We measured a typical 1$\sigma$ uncertainty ranging from 0.15 km~s$^{-1}$ in the case of M 15, to 0.8 km~s$^{-1}$ for NGC 5927. 
On the contrary, PA$_{0}$ is more sensitive to 
the spatial distribution of the adopted sample, with an associated uncertainty 
at the $\pm$ 30$^{\circ}$ level in the best cases.

The considered clusters span a wide range of rotation amplitude, 
from no rotation within the uncertainties (NGC 6752) to an amplitude of more than 3.5 km~s$^{-1}$ (NGC 2808 and M 15). 
We note that the two clusters with clear rotation pattern,  NGC 2808 and M 15, are among
the most peculiar clusters in terms of multiple populations, with an extended horizontal branch morphology
(see for a recent review \citealp{gratton12} and references therein).
For the six clusters already considered in previous studies 
(i.e., all the sample clusters but NGC 5927),
we confirm the results reported in the literature, while we were able to
detect for the first time a significant amplitude of rotation for 
the metal-rich cluster\footnote{The value tabulated in the 
\citealp{harris96} catalogue for NGC 5927 is [Fe/H]= --0.49 dex; it was
obtained by averaging the [Fe/H] derived by \citealp{armandroff88,francois91,carmet}.} NGC 5927, 
A$_{\rm{rot}}$ =  2.6 km~s$^{-1}$.
  
\begin{figure}
   \centering
   \includegraphics[width=\columnwidth]{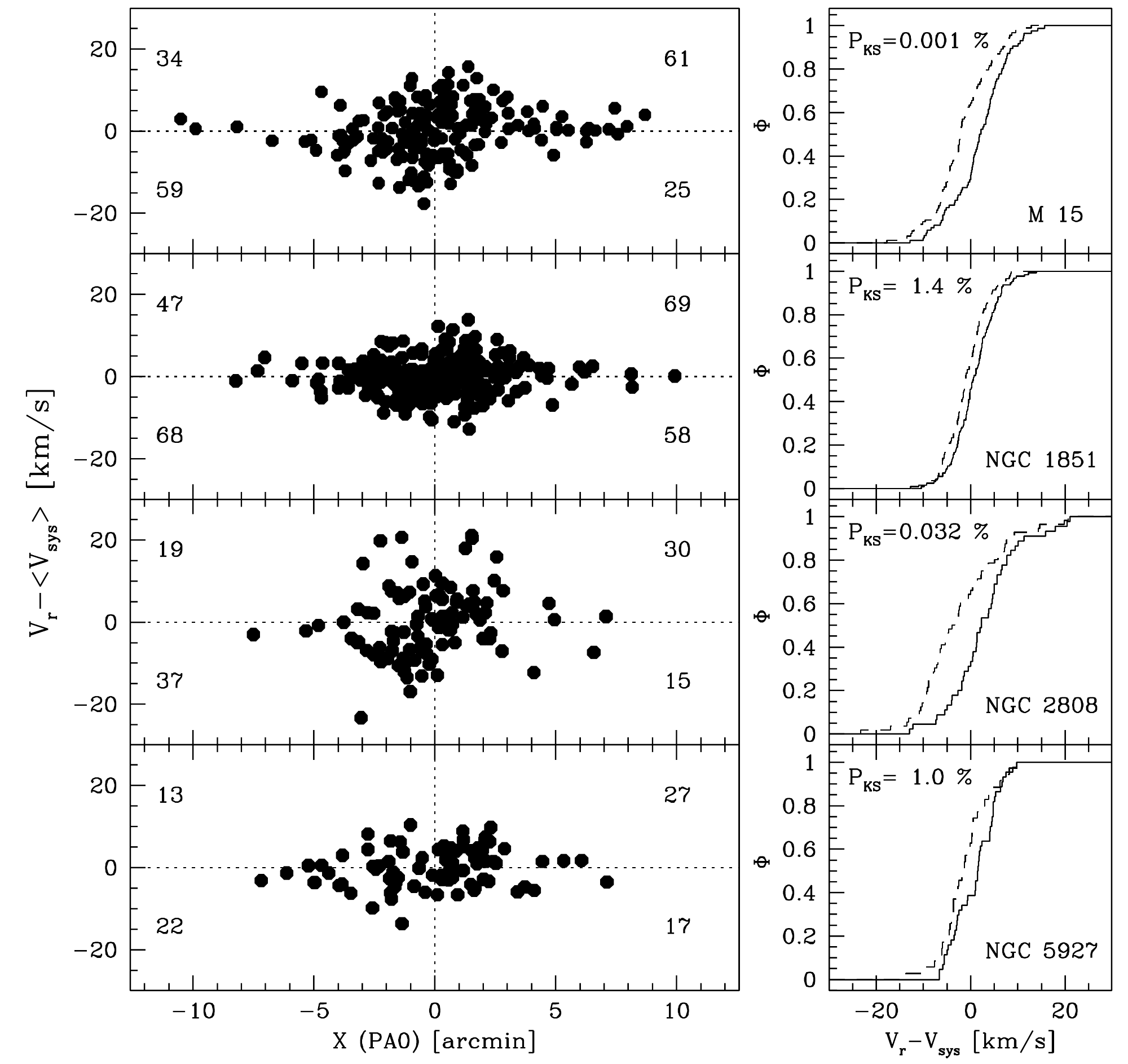}
   \caption{Rotation curves for M 15, NGC 1851, NGC 2808, and NGC 5927. 
    {\em Left panels}: V$_{\rm{r}}$ in the system of the cluster as a function 
    of distance from the centre projected onto the axis perpendicular 
    to the best-fit rotation axis found in Fig.~\ref{ROTAZIONI}. The number of stars in each quadrant is also
    shown. {\em Right panels}: comparison of the cumulative V$_{\rm{r}}$ 
    distributions of stars with X(PA0) $>$ 0.0 (continuous lines) and X(PA0) $<$ 0.0 (dashed lines). 
    The probability that the two distributions are drawn from the same parent population 
    (according to a KS test) is reported in each panel. 
    We show rotation curves only for the four clusters with P$_{\rm{KS}}$ $<$ 2.5\%.}
              \label{CURVAX}
    \end{figure}

In Fig.~\ref{CURVAX} we show the rotation curves for M 15, NGC 1851, NGC 2808, and NGC 5927;
these are the four clusters for which significant rotation was detected.
In the right-hand panels, the V$_{\rm{r}}$ distribution of stars lying on opposite sides with respect 
to the rotation axis are compared. If the clusters  were non-rotating, the two distributions would be 
identical, while a shift should be apparent with significant rotation \footnote{We note that 
the degree to which the two distributions differ also depends on the ratio between rotation 
and velocity dispersion and on the actual shape of the rotation curve.}.
A Kolgomorov-Smirnov test shows instead that it is relatively unlikely that the observed 
patterns may emerge by chance from non-rotating systems (see left-hand panels of Fig.~\ref{CURVAX}).

\section{Trends with cluster parameters}\label{parametri}
B12 used kinematic data for several GCs to explore
the dependences of several GC parameters on the A$_{\rm{rot}}$ and A$_{\rm{rot}}$/$\sigma_{0}$.
In particular, these authors made use of the large database ($\simeq$ 2000 stars) 
collected in the framework of the Na-O anti-correlation and HB program 
(see for example \citealp{carrettaUVES,carrettaGIRAFFE} for a more detailed description). 
The B12 database included 24 GCs that partially overlap with our sample 
(see also \citealp{meylan97}), and our study is largely homogeneous with their analysis.
Therefore, we added three new clusters to the compilation in B12 (i.e., NGC 4372, NGC 4833, and NGC 5927) and 
considered for the clusters in common our own values of the central velocity dispersion and A$_{\rm{rot}}$.

Table~\ref{AMMASSI} lists $\sigma_{0}$ and A$_{\rm{rot}}$ estimates for all the clusters,
together with other relevant parameters from various sources.\begin{table*}
\centering
 \caption{Cluster parameters}
 \renewcommand{\tabcolsep}{0.23cm}
 \label{AMMASSI}

\begin{tabular}{l c c c c c c c c c c  }
            \hline\hline
Cluster   & $\sigma_{0}$  & $\epsilon_{\sigma}$ &  A$_{\rm{rot}}$  &  $\epsilon _{\rm{A}}$ &  [Fe/H] &    HBR   &  M$_{\rm{V}}$  & ell &  $\log$ $\rho_{0}$     & R$_{\rm{G}}$ \\          
          & km~s$^{-1}$ & km~s$^{-1}$       & km~s$^{-1}$    & km~s$^{-1}$         & dex     &          &                &     &  L$_{\sun}$ pc$^{-3}$  & kpc          \\
\hline
NGC 104   &  9.6   &  0.6  &   4.4  &   0.4   &  -0.76  & -0.99  & -9.42  &  0.09  &   4.88 &   7.4 \\   
NGC 288   &  2.7   &  0.8  &   0.5  &   0.3   &  -1.32  &  0.98  & -6.75  & 0.00   & 1.78   & 12.0 \\        
NGC 1851  &  12.3\tablefootmark{a}   & 1.5   &  1.6\tablefootmark{a}    &   0.5    & -1.16  &  -0.32 &  -8.33 &  0.05  &  5.09  &  16.6  \\         
NGC 1904  &  5.3   &  0.4  &   0.6  &   0.5   &  -1.58  &  0.89  & -7.86  & 0.01   & 4.08   & 18.8   \\ 
NGC 2808  & 18.8\tablefootmark{a}    & 4.0   &  4.7\tablefootmark{a}    &   0.2  &   -1.18  &  -0.49 &  -9.39 &  0.12  &  4.66  &  11.1 \\  
NGC 3201  &  4.5   &  0.5  &   1.2  &   0.3   &  -1.51  &  0.08  & -7.45  & 0.12   & 2.71   & 8.8  \\ 
NGC 4372  &  4.9\tablefootmark{a}    &  1.2  &   1.0\tablefootmark{a}   &   0.5  &     -2.17\tablefootmark{b}  &   1.00\tablefootmark{c} &  -7.77\tablefootmark{c} &  0.15\tablefootmark{c}  &  2.06\tablefootmark{b} &  7.1\tablefootmark{c} \\
NGC 4590  &  2.4   &  0.9  &   1.2  &   0.4   &  -2.27  &  0.17  & -7.37  & 0.05   & 2.57   & 10.2  \\ 
NGC 4833  &  5.5\tablefootmark{a}    &  1.5  &   1.2\tablefootmark{a}   &  0.4   &   -1.85\tablefootmark{b} &     0.93\tablefootmark{c} &  -8.16\tablefootmark{c} &  0.07\tablefootmark{c}  &  3.00\tablefootmark{b}  &  7.0\tablefootmark{c} \\
NGC 5024  &  4.4   &  0.9  &   0.0  &   0.5   &  -2.06  &  0.81  & -8.71  & 0.01   & 3.07   & 18.4   \\ 
NGC 5139  &   19.0 &   1.0 &    6.0 &   1.0  &   -1.64  &   $$-$$&  -10.26&   0.17 &   3.15 &      6.4 \\  
NGC 5904  &  7.5   &  1.0  &   2.6  &   0.5   &  -1.33  &  0.31  & -8.81  & 0.14   & 3.88   & 6.2   \\ 
NGC 5927  & 11.0\tablefootmark{a}    &  2.0  &  2.6\tablefootmark{a}    &  0.8  &     -0.49\tablefootmark{b}  &  -1.00\tablefootmark{c} &  -7.80\tablefootmark{c} &  0.04\tablefootmark{c}  &  4.09\tablefootmark{b}  &  7.3\tablefootmark{c} \\
NGC 6121  &  3.9   &  0.7  &   1.8  &   0.2   &  -1.18  & -0.06  & -7.19  & 0.00   & 3.64   & 5.9     \\ 
NGC 6171  &   4.1  &   0.3 &    2.9 &   1.0  &   -1.03  &  -0.73 &  -7.12 &  0.02  &  3.08  &  3.3  \\  
NGC 6218  &  4.7   &  0.9  &   0.3  &   0.2   &  -1.33  &  0.97  & -7.31  & 0.04   & 3.23   & 4.5     \\ 
NGC 6254  &   6.6  &   0.8 &    0.4 &   0.5  &   -1.57  &   0.98 &  -7.48 &  0.00  &  3.54  &  4.6   \\  
NGC 6388  &  18.9  &  0.8  &   3.9  &   1.0   &  -0.45  & -0.65  & -9.41  & 0.01   & 5.37   & 3.1     \\ 
NGC 6397  &  4.5   &  0.6  &   0.2  &   0.5   &  -1.99  &  0.98  & -6.64  & 0.07   & 5.76   & 6.0     \\ 
NGC 6441  &  18.0  &  0.2  &   12.9 &   2.0   &  -0.44  & -0.76  & -9.63  & 0.02   & 5.26   & 3.9     \\ 
NGC 6656  &  6.8   &  0.6  &   1.5  &   0.4   &  -1.70  &  0.91  & -8.50  & 0.14   & 3.63   & 4.9     \\ 
NGC 6715  &  16.4  &  0.4  &   2.0  &   0.5   &  -1.56  &  0.54  & -9.98  & 0.06   & 4.69   & 18.9    \\ 
NGC 6752  &  8.2\tablefootmark{a}    &  0.6  &   0.7\tablefootmark{a}   &   0.2   &  -1.55    &  1.00  & -7.73  & 0.04   & 5.04   & 5.2   \\ 
NGC 6809  &  2.7   &  0.5  &   0.5  &   0.2   &  -1.93  &  0.87  & -7.57  & 0.02   & 2.22   & 3.9    \\ 
NGC 6838  &  2.3   &  0.2  &   1.3  &   0.5   &  -0.82  & -1.00  & -5.61  & 0.00   & 2.83   & 6.7    \\ 
NGC 7078  &  13.2\tablefootmark{a}   &  1.5  &   3.6\tablefootmark{a}   &   0.1   &  -2.33    &  0.67  & -9.19  & 0.05   & 5.05   & 10.4    \\ 
NGC 7099  &  5.0   &  0.9  &   0.0  &   0.0   &  -2.33  &  0.89  & -7.45  & 0.01   & 5.01   & 7.1     \\ 
\hline

\end{tabular}
  
 \tablefoot{
All parameters are reported from \citet{bellazzini12} except:\\
\tablefoottext{a}{this work}\\
\tablefoottext{b}{\citealp{harris96} (2010 edition)}\\
\tablefoottext{c}{\citet{mackey05}}\\
Meaning of columns:\\ (1) Cluster name; 
 (2) central radial velocity dispersion; (3)  error on $\sigma_{0}$;
 (4) projected rotation amplitude; (5) error on A$_{\rm{rot}}$;
 (6) mean iron abundance ratio; (7) HB morphology, where HBR = (B -- R)/(B + V + R), 
 where B is the number of stars bluer than the instability strip, R redder, and 
 V the number of variables in the strip; (8) the integrated $V$ magnitude;
 (9) the isophotal ellipticity $\epsilon$ = 1 -- (b/a); (10) central luminosity density;
 (11) distance from the Galactic centre (kpc).
}
\end{table*}
\begin{figure}
   \centering
   \includegraphics[width=\columnwidth]{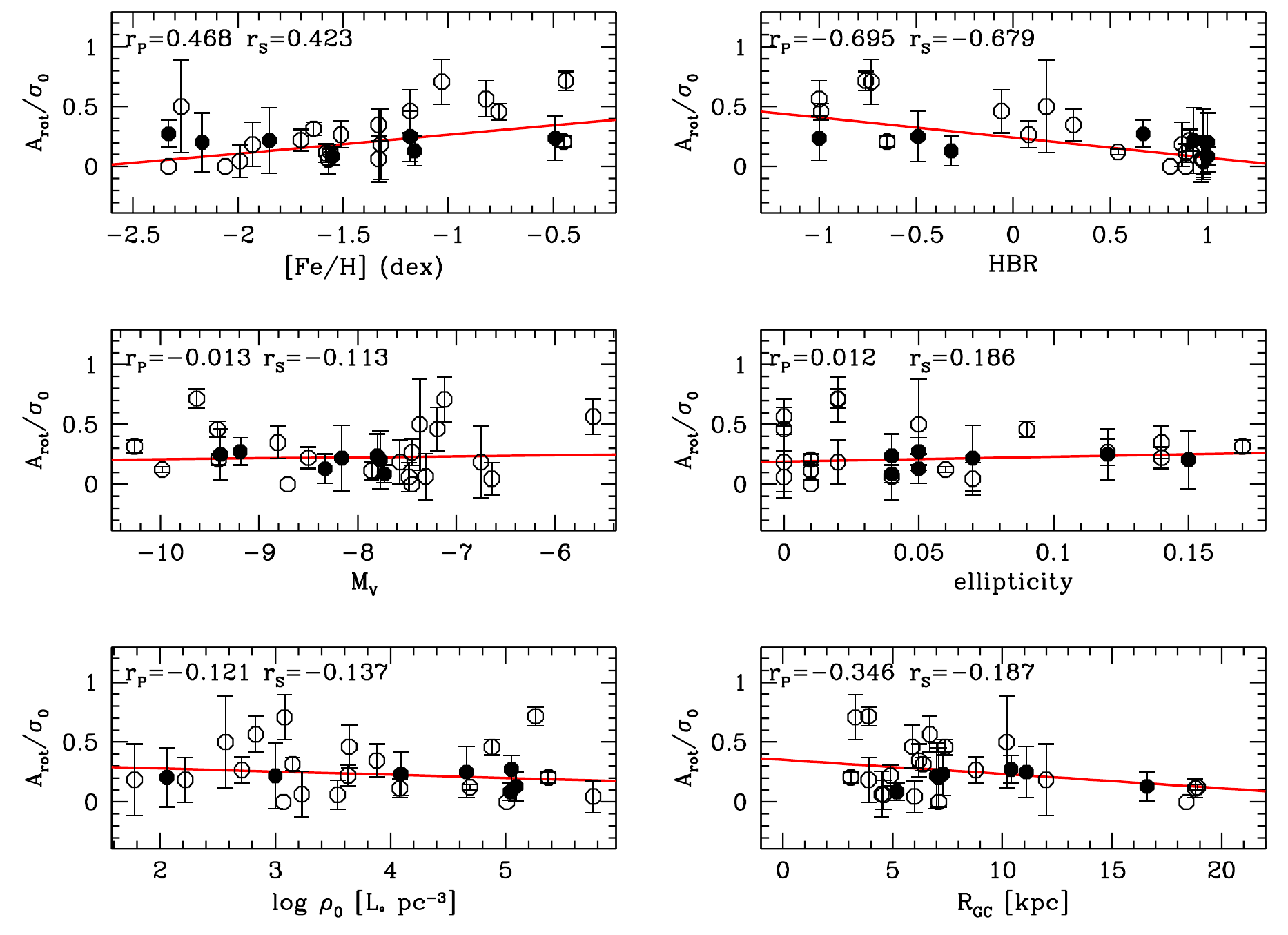}
   \caption{Ratio between the amplitude of the rotation A$_{\rm{rot}}$ and the central velocity 
   dispersion $\sigma_{0}$  versus various other parameters.
   Red lines mark weighted linear fits to the clusters, and the correlation 
   coefficients are reported at the top of each panel: r$_{\rm{S}}$ stands for the Spearman and 
   r$_{\rm{P}}$ for the Pearson coefficient. Empty circles are data from B12, while filled circles are 
   our own estimates.}

              \label{TREND}
    \end{figure}
In Fig.~\ref{TREND} we show the behaviour of  the ratio 
A$_{\rm{rot}}$/$\sigma_{0}$ as a function of
metallicity, the HB morphology parameter HBR = (B -- R)/(B + V + R)
(\citealp{lee90}, see caption in Table~\ref{AMMASSI} for its definition), 
the absolute integrated V magnitude (M$_{\rm{V}}$),  
the logarithm of the central luminosity density ($\log$ $\rho_{0}$),
and the distance from the Galactic centre (R$_{\rm{GC}}$). 
The same figure also reports the Pearson (r$_{\rm{P}}$) and Spearman (r$_{\rm{S}}$)
correlation coefficients. The ratio A$_{\rm{rot}}$/$\sigma_{0}$  does not show any clear correlation 
with M$_{\rm{V}}$, ellipticity, $\log$ $\rho_{0}$, and R$_{\rm{GC}}$. 
On the contrary, a clear correlation emerges between A$_{\rm{rot}}$/$\sigma_{0}$ 
with [Fe/H] and HBR (see B12).
For more metal-rich clusters the relevance of ordered motions with respect to pressure is stronger.
According to a two-tailed Student's test, the probability that a Spearman 
rank correlation coefficient equal to or higher than the observed one 
(r$_{\rm{S}}$ =  0.423) is produced by chance from uncorrelated quantities is 
P$_{\rm{t}}$ =  3.0\% (27 clusters), so the correlation can be considered as 
statistically significant.
In addition, the A$_{\rm{rot}}$/$\sigma_{0}$ ratio appears to be significantly correlated with 
the HB morphology (P$_{\rm{t}}$ = 1 $\times$ 10$^{-4}$) in the sense that 
clusters with redder HB have greater fractions of ordered motions with respect to pressure support. 

\begin{figure}
   \centering
   \includegraphics[width=\columnwidth]{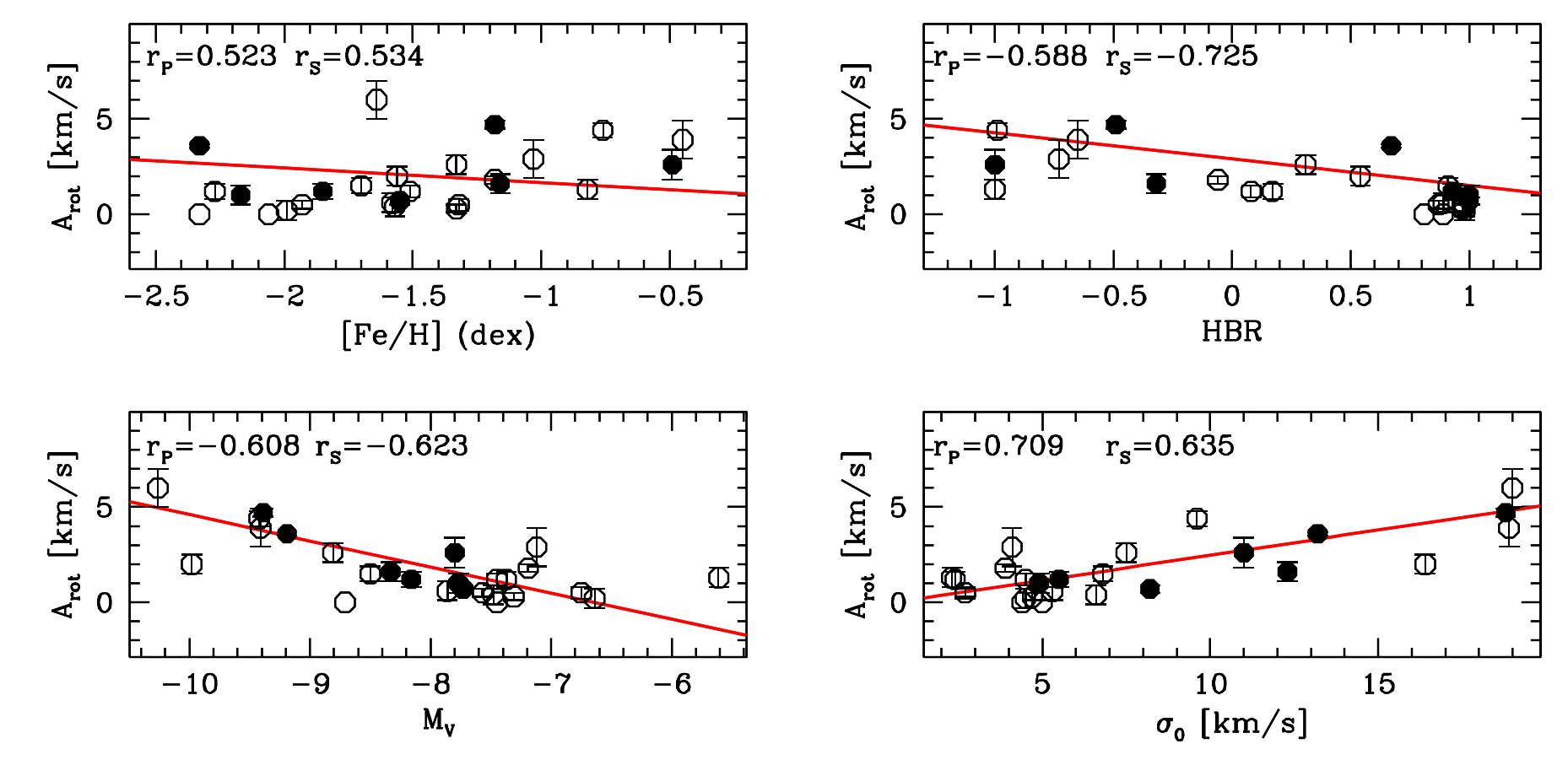}
   \caption{Run of the amplitude of the rotation A$_{\rm{rot}}$ vs. 
   versus various other parameters.
   Red lines mark weighted  linear fits to the clusters (filled and empty) and the correlation 
   coefficients are reported at the top of each panel: r$_{\rm{S}}$ stands for the Spearman and 
   r$_{\rm{P}}$ for the Pearson coefficient. Empty circles are from B12, while filled circles are 
   our own estimates.}

              \label{TRENDROT}
    \end{figure}

Additionally, Fig.~\ref{TRENDROT} shows that A$_{\rm{rot}}$ has statistically significant correlation with 
HBR (P$_{\rm{t}}$ = 1 $\times$ 10$^{-5}$), M$_{\rm{V}}$ (P$_{\rm{t}}$ = 5 $\times$ 10$^{-4}$), 
$\sigma_{0}$(P$_{\rm{t}}$ = 2 $\times$ 10$^{-4}$), and [Fe/H] (P$_{\rm{t}}$ = 
4 $\times$ 10$^{-3}$)\footnote{We caution, however, that the statistics quoted for P$_{\rm{t}}$ could be slightly 
misleading because a correlation may emerge even in a random dataset, whereas there are enough parameters and 
enough correlation plots.}.
All the above results agree well with those reported by B12.

\section{Summary and conclusions}\label{Conclusioni}

We used the radial velocity estimates obtained from the second internal data release of data products to ESO of the Gaia-ESO survey to study the kinematics of seven Galactic GCs.
We confirm the central velocity estimates reported in the literature for 
NGC 1851, M 15, NGC 4372, and NGC 4833, while we found that there is a real discrepancy between
the central dispersion from radial velocities and that from proper motions for NGC~6752.
For NGC 2808, our sample is too sparse to draw useful conclusions about $\sigma_0$.
Finally, we provided for the first time a velocity dispersion profile and a central velocity dispersion estimate for NGC 5927, albeit uncertain (see Sect.~\ref{sigma}).
We searched for systemic rotation in all the studied clusters and found 
significant rotation patterns  (A$_{\rm{rot}} \geq$ 2.5 km~s$^{-1}$) in NGC 2808, 
NGC 5927, and M 15 and a marginal detection for NGC 1851 (see Sect.~\ref{rotazioni}).

We demonstrated that the radial velocities delivered from the Gaia-ESO survey pipeline
have  sufficient quality to be used in a profitable way in a kinematic study and made 
public a large database of radial velocities of GCs members for future research.
For example, we verified that the uncertainties on individual radial velocity estimates from the
survey pipeline are fully reliable because they match the errors
on the mean derived from multiple independent measures.

When all the archival data will be incorporated into the Gaia-ESO survey and abundances will be available for all the analysed stars, the final large  dataset will permit insightful analyses of the internal motions of the clusters.
For example, it will allow us to correlate the presence and amplitude of rotation 
with the cluster parameters, different chemistry and/or sub-population.
Moreover, the Gaia satellite will provide 3D 
kinematical data for a significant number of these stars (see \citealp{pancinoGAIA}), so that the analysis we presented here can be 
considered as a preparatory study aimed at a complete exploitation of the Gaia data.

\begin{acknowledgements}
We thank the referee, N.~Martin, for the careful reading of the manuscript and for the useful comments and suggestions that 
helped to improve the quality of the paper significantly.
We acknowledge the support from INAF and Ministero dell' Istruzione, dell' Universit\`a e della Ricerca (MIUR) 
in the form of the grants ``Premiale VLT 2012'' and  
``The Chemical and Dynamical Evolution of the Milky Way and Local Group Galaxies'' (prot. 2010LY5N2T).
PdL and ARB acknowledge the support of French Agence Nationale de la Recherche, under
contract ANR-2010-BLAN-0508-01OTP, and the Programme National de Cosmologie et Galaxies.
This work was partly supported by the European Union FP7 programme through ERC grant number 
320360 and by the Leverhulme Trust through grant RPG-2012-541.
The results presented here benefit from discussions held during the Gaia-ESO workshops and 
conferences supported by the ESF (European Science Foundation) through the GREAT Research Network Programme.
This research has made extensive use of NASA's Astrophysics Data
System Bibliographic Services, and of the SIMBAD database
and VizieR catalogue access tool, CDS, Strasbourg, France.  
\end{acknowledgements} 

\bibliographystyle{aa}
\bibliography{bibliography}

\end{document}